\pdfoutput=1
\documentclass[prl,twocolumn,showpacs]{revtex4-1}
\usepackage{graphicx}
\usepackage{amsmath}
\usepackage{amssymb}
\usepackage{dcolumn}
\usepackage{natbib}
\usepackage{latexsym}
\usepackage{rotating}
\usepackage{color}
\usepackage{latexsym}
\usepackage{subfigure}
\usepackage{float}
\usepackage{epsfig}
\usepackage{psfrag}
\usepackage{bm}
\usepackage{amsthm}
\usepackage{eucal}
\usepackage{mathrsfs}
\usepackage{braket}
\usepackage{psfrag}
\usepackage{enumerate}
\usepackage{longtable}
\usepackage{subfigure}
\usepackage{bm}
\usepackage{hyperref}
\hypersetup{
colorlinks=true,final=true,
        linkcolor=red,
        citecolor=blue,
        filecolor=blue,
        urlcolor=blue,
}

\usepackage{dcolumn}
\usepackage{bm}
\usepackage{dcolumn}
\usepackage{bm}

\begin{document}

\draft

\title{Fermiology in a Local Quantum Critical Metal}. 
\author{Mukul S. Laad$^1$, S. Koley$^2$ and A. Taraphder$^2$}

\affiliation{$^{1}$ Institute of Mathematical Sciences, Taramani, Chennai 600113, India \\
$^{2}$Department of Physics and Centre for Theoretical studies,\\
Indian Institute of Technology, Kharagpur 721302 India}

\date\today

\begin{abstract}
Recent experimental work has brought the twin issues of the origin of non-Lifshitz-Kosevich
scaling in de Haas van Alphen (dHvA) and  its precise relation to anomalously broad
non-quasiparticle spectral features in ``strange" metals, to the forefront. Here, we revisit
these issues in the specific context of a ``local" quantum critical phase in an extended
periodic Anderson model (EPAM).  In contrast to the famed Kondo-RKKY scenarios for local quantum criticality, strong local valence fluctuations cause Kondo destruction in the EPAM.  We uncover a common underlying element, namely, the Kondo-destruction-driven infra-red continuum branch-cut behavior in the one-electron propagator,  as the relevant feature that governs both non-Lifshitz-Kosevich scaling in dHvA and anomalously broad non-quasiparticle spectral responses in such a ``strange" metal.  Employing a non-perturbative scheme to treat effects of non-magnetic disorder in this version of a local strange metal, we propose a modified Dingle scaling that can also be used to test ``local'' criticality scenarios.  Thus, our findings potentially afford an internally consistent description of novel fermiology expected to manifest in strange metals arising as a result of Kondo-destruction coming from 
an underlying orbital-selective Mott transition.

\end{abstract}

\maketitle

Recent finding of Fermi surface reconstruction (FSR) and, in particular, of small hole pockets in 
underdoped cuprates (also in increasing number of other systems~\cite{gil}) in high-field de 
Haas-van Alphen data has spurred intense theoretical activity that goes back to the very 
foundations of fermiology in correlated electronic systems: Is the Luttinger theorem, a 
cornerstone of interacting Fermi liquids in metals, valid in its original formulation?  If it is 
not, have strong correlations non-perturbatively restructured the ``large'' Fermi surface (FS) of 
the interacting Fermi liquid (FL)?  Is this restructuring due to conventional order, or is 
it a more exotic manifestation of the non-perturbative extinction of the Landau-FL metallic state 
itself?

Intrigued by these developments, several theoretical studies have insightfully addressed these 
issues. These range from (i) a demonstration of the non-perturbative nature of the Luttinger 
theorem~\cite{oshikawa}, (ii) a topological quantum phase transition that, in its varied 
guises, involves penetration of Green function zeros to the poles, obliterating the adiabatic 
continuity central to FL theory.  Concrete studies~\cite{rmft,civelli,imada,laad1} associate this
feature explicitly with ``strange" metallicity and breakdown of the FL picture. Still, how to 
precisely characterise the critical FS, poised on the boundary between the Luttinger-FS of the 
heavy Fermi liquid (HFL) and the reconstructed FS of the ordered phase(s), is an intriguing 
and subtle issue that deserves more careful attention.

Traditionally, quantum oscillation studies have played a crucial role in fermiology by 
comprehensively mapping out complex multi-sheeted (pocketed) FS of carriers in metals, 
including strongly correlated $d$ and $f$ band systems~\cite{julian}. As long as the 
FL quasiparticle description holds, dHvA results can be cleanly interpreted in terms of the 
traditional Lifshitz-Kosevich (LK) theory. However, how they should be interpreted in 
non-FL metals, where the Landau quasiparticle structure in the fermion 
propagators breaks down, was always~\cite{stamp} enigmatic, albeit of obvious interest in the 
context of the strange normal state of HTSC cuprates. Recent observation of anomalous 
$T$-dependence (i.e, non-Lifshitz-Kosevich (LK) form) of de Haas-van Alphen (dHvA) amplitudes in 
CeCoIn$_{5}$~\cite{julian} has given a fresh boost to this issue.  

Complementarily, tremendous strides in angle resolved photoemission (ARPES) technology now allow a
truly unprecedented mapping of renormalised band dispersions {\it and} lineshapes in
correlated systems, including the strange metal~\cite{laser-arpes,utpal}. While dHvA thus measures
the FS (ground state property), ARPES measures the actual one-fermion excitation spectrum: in 
principle, both result from a one-electron propagator, $G({\bf k},\omega)$, which encodes all 
interaction effects via an irreducible one-fermion self-energy, $\Sigma({\bf k},\omega)$. In this 
context, serious difficulties show up when one attempts to correlate ARPES with dHvA results: even
in $f$-electron metals for which most dHvA data exist, {\it apparent} accord with the LK form 
is routinely found, while complementary spectral responses show strong violation of the 
Landau FL theory.  Since traditional LK scaling itself is crucially based upon the analytic 
structure of the Landau FL theory, this immediately leads to an internal inconsistency when one 
attempts to understand fermiology in a unified way.  In cuprates, admittedly the best example of 
a non-FL metal in $D>1$, the situation is worse for want of dHvA data in strange metals. 
Theoretically, there is thus a great need to have well-controlled scenario(s) where clean 
correlations between dHvA and ARPES can be drawn, and both correlated with anomalous transport.  
This is a highly non-trivial task for correlated metals, and it is only possible to make precise 
theoretical statements in very few well-controlled limits: e.g, large-$N$ or large-$D$: DMFT and 
its cluster extensions are the methods of choice in cases where Mottness is implicated in 
strange metallicity.  Moreover, they are a particularly well-suited choice in the case of 
``local quantum critical" behavior. 

Motivated thus, as well as our studies using dynamical mean field theory (DMFT) for an extended 
periodic Anderson model (EPAM)~\cite{laad1,swagata}, we seek to correlate dHvA and ARPES across local 
Lifshitz quantum criticality associated with an orbital-selective Mott transition (OSMT). 
We restrict ourselves to symmetry unbroken states: additional FS reconstruction(s) must 
occur upon onset of (un)conventional quasiclassical order, a topic we leave for future study.  In this paper, we use parameters from our earlier DMFT study on the EPAM~\cite{laad1} to present numerical data for renormalized Fermi (Luttinger) surfaces, ``band'' structures and ARPES lineshapes across the OSMT.   

\noindent {\bf de-Haas van Alphen Effect}:  Recently, an exciting proposal for dHvA effect in a quantum 
critical metal, based upon a non-perturbative holographic correspondence, was made~\cite{hartnoll}. 
The main result is that the dHvA magnetic oscillations in a class of (2+1) dimensional quantum critical 
theories are anomalous in a very specific sense:

\begin{equation}
\chi_{osc}=A\frac{\pi Tck_{F}^{4}}{eB^{3}} cos(\frac{\pi ck_{F}^{2}}{eB})\sum_{n=0}^{\infty}e^{-(cT\mu/eBk_{F}^{2})(T/\mu)^{2\nu-1}F_{n}(\nu)}
\end{equation}

\noindent where $F_{n}(\nu)$ is a function of $\omega/T$.  Remarkably, 
the standard LK formula is recovered for FL and marginal-FL choices, where
$F_{n}(\nu=1/2)\simeq (n+1/2)$ and the sum over $n$ can be explicitly done.  
However, for $\nu<1/2$, at ``high" $T$ (while $T<<\mu$ still), the decay of the 
dHvA amplitudes deviates from the textbook ($e^{-bT}$) LK form, yielding
$\chi_{osc}\simeq$ exp$(-T^{2\nu})$.  The essential point is that this can 
be represented as standard LK scaling {\it if} the effective quasiparticle
mass were chosen as $m^{*}\simeq (k_{F}^{2}/\mu)(\mu/T)^{1-2\nu}$, infra-red 
divergent if $\nu<1/2$. 

We now show that exactly the same form also directly obtains as a 
result of the lattice orthogonality catastrophe in the EPAM in DMFT.  
Specifically, for a range of values of $U_{fc}^{(1)}<U_{fc}<U_{fc}^{(2)}$ in
the EPAM, we have shown earlier~\cite{laad1} that DMFT yields a selective metal: one (say $`a'$-band) 
band is selectively Mott-gapped at low energies, while the other (say $`b'$-band) shows 
precisely the infra-red singular behavior, $G_{bb}(\omega)\simeq |\omega|^{-(1-\alpha)}$
~\cite{laad1}, characteristic of strange metallicity, as in hidden-FLT~\cite{pwa}. 
The upshot is that the corresponding self-energy now behaves like 
Im$\Sigma_{b}(\omega)\simeq$ Im$G_{bb}^{-1}(\omega)\simeq |\omega|^{1-\alpha}$ near $\omega=\mu$.
The fermion spectral function exhibits an anomalous broad peak with a ($T=0$) band dispersion of 
the form $e^{i\theta}\omega^{1-\alpha}=k-k_{F}$ near $k_{F}$ (see below), the precise form 
{\it necessary} to get non-LK scaling in the holographic theory.  This peak is obviously not an 
FL quasiparticle: viewed as a pole, its weight vanishes like a power law; i.e,
as $z(\omega)\simeq \omega^{\alpha}$ as one hits the real axis at $k=k_{F}$.  
The singular locus of Im$G_{bb}(k,\omega)$ is thus a branch-cut emanating from 
$\omega=\mu$.  The corresponding effective mass diverges in the infra-red as
(reintroducing the Fermi energy $E_{F}$ as an effective high-energy cut-off)
$m^{*}\simeq (k_{F}^{2}/\mu)(\mu/\omega)^{\alpha}$.  In the local critical 
metal, with $\omega/T$ scaling of $G_{bb}(\omega)$ at finite $T$, one may replace 
$T$ for $\omega$, yielding $m^{*}\simeq (k_{F}^{2}/\mu)(\mu/T)^{\alpha}$. Finally, 
identiying $\alpha$ with $(1-2\nu)$ in the holographic theory yields Eq.(1) {\it if} we now use 
this effective dynamical mass (diverging at the OSMT) in the {\it traditional} LK formula 
(where $m^{*}$ is constant at low $T$).  

That exactly the same result can be derived within the structure of more traditional condensed 
matter theory can be seen as follows.  Following Hartnoll {\it et al.}, we start with the free
energy, $\Omega=\frac{eBAT}{\pi c}Re\sum_{n=0}^{\infty}\sum_{k=1}^{\infty}\frac{1}{k}e^{i2\pi kl_{*}(n)}$, where $l_{*}(n)$ are defined as solutions to $F(\omega_{n},l_{*}(n))=0$, with 
$F=0$ defining the singular locus of the fermion Green function in a magnetic field~\cite{hartnoll}. 
Then we note that $\omega_{n}=2\pi iT(n+1/2)$, and use $k^{2}=\frac{2leB}{c}$. Now, solving 
for $l_{*}(n)$ from $F(\omega_{n},l_{*}(n))=0$, inserting into $\Omega$ above, and 
differentiating twice gives Eq.(1), as done by Hartnoll {\it et al}. Apparently, we have 
only {\it needed} the anomalous form of $e^{i\theta}\omega^{1-\alpha}=k-k_{F}$ and the eqn. 
for $\Omega$ to get Eq.(1). It thus follows that the non-LK scaling is rationalisable as a direct 
consequence of the ``inverse'' lattice orthogonality catastrophe that accompanies onset of 
the selective-Mott state in DMFT (our detour is thus {\it not} linked to AdS-CFT ideas, 
though they seem to be related~\cite{sachdev}). In the HFL phase of the EPAM for $U_{fc}< 
U_{fc}^{(1)}$~\cite{laad1}, a small but finite $z (\omega=E_{F})$ ensures, by adiabatic 
continuity to a free Fermi system, that the traditional LK scaling must hold below a severely 
renormalised lattice coherence scale, $E^{*}=z_{FL}\mu$.  

One might wonder about the fate of the cyclotron orbits in the dHvA effect in the strange metal. 
The destruction of Landau quasiparticles that accompanies the divergent effective mass at the 
orbital selective metal criticality would seem to pre-empt the conventional description, where 
$\omega_{H}=\omega_{c}=(eB/m^{*})$. Though one must certainly be able to connect this picture with
the heavy-FL regime, thanks to its adiabatic continuity with the free Fermi system, one certainly 
requires that it be invalid in the strange metal. This is because lack of Landau quasiparticles 
necessarily implies breakdown of the conditions under which any quasiclassical band approach 
is valid.  A generalization of the conventional treatment is thus in order, but a strong 
constraint is that it must be general enough to reduce to the conventional picture above when 
Landau quasiparticles obtain. We now detail a generalized approach that fulfils the above 
constraint. The most general way to a (generalized) cyclotron frequency, regardless of whether or 
not Landau quasiparticles exist, is to notice~\cite{shastry} that it ($\omega_{H}$) can be 
extracted from the high-frequency asymptotic form of the full conductivity tensor in a magnetic 
field. From the Kubo formula with $a,b=x,y,\, \sigma_{ab}(\omega)=Q_{ab}(\omega)/(-i\omega)$ where 
 
\begin{equation}
Q_{ab}(t)=\epsilon_{0}\omega_{p}^{2}\delta_{ab}\delta(t)-i\langle[J_{a}(t),J_{b}(0)]\rangle\theta(t)
\end{equation}

\noindent This leads to dominant short-time behaviors, $Q_{xx}(t)\simeq \epsilon_{0}\omega_{p}^{2}\delta(t)$ 
and $Q_{xy}(t)\simeq -i\langle[J_{a}(0),J_{b}(0)]\rangle\theta(t)$ with $t\rightarrow 0$. Hence, $\sigma_{xx}
(\omega)=\epsilon_{0}\omega_{p}^{2}/(-i\omega)$ and $\sigma_{xy}(\omega)=i\langle[J_{x},J_{y}]\rangle/\omega^{2}$ 
as $\omega\rightarrow\infty$.  Taking the ratio, one finds that tan$\theta_{H}(\omega)=\omega_{H}/(-i\omega)$ in 
this limit, where

\begin{equation}
\omega_{H}=\frac{i\langle[J_{x},J_{y}]\rangle}{\epsilon_{0}\omega_{p}^{2}}
\end{equation}

\noindent is precisely the required generalization of the cyclotron orbit frequency we seek. It 
correctly reduces to the required answer in the Landau FL case, since (for the simplest parabolic 
band dispersion) $i\langle[J_{x},J_{y}]\rangle =\omega_{c}(\epsilon_{0}\omega_{p}^{2})$, leading 
to $\omega_{H}=\omega_{c}$. But this can also be readily evaluated in the strange metal regime, 
since the denominator is just the celebrated $f$-sum rule for the optical
conductivity, and, in our specific case, this can be directly evaluated from our earlier DMFT results. 
The numerator can also be explicitly evaluated for any lattice model such as ours, the treatment 
closely paralleling that of Shastry {\it et al.}  Explicitly, the commutator in the high-$T$ 
(incoherent metal limit for the Hubbard model) is Lim$_{B\rightarrow 0}\frac{1}{iB}
\langle[J_{x},J_{y}]\rangle = -t^{4}\beta^{2}Nx(1-x)[x^{2}-(1-x)^{2}/4] +O(\beta^{4})$, 
with $\beta=1/k_{B}T$ and $x=(1-n)$ the hole density per site. We expect a similar estimate for 
our EPAM. Actually, at lower $T$, this commutator is considerably more complicated and 
involves coupled, short-ranged charge and spin correlations: it remains an open and intriguing issue 
whether this thus hides the spinon backflow caused by accelerating the actual 
electron by a magnetic field in a Tomonaga-Luttinger-liquid-like description~\cite{pwa}. The closest 
we can attempt to qualitatively connect to this picture is by identifying the unquenched local-moment 
subsystem in the (paramagnetic) strange metal with the spinons in RVB theory: in this context, it is 
interesting that it is precisely the ``inverse orthogonality catastrophe'' arising from strong 
local correlations that is at the heart of the ``hidden Fermi liquid'' theory of Anderson.   
In the local critical metal we find, the high-$T$ incoherent metal persists down to low $T$, since 
orbital-selective Mottness in the EPAM implies existence of a partially unquenched and critically fluctuating local-moment regime resulting
from destruction of the local Kondo screening down to low $T$ in absence of 
conventional symmetry breaking, at least within DMFT. Plugging the commutator $[J_{x},J_{y}]$ and 
using the $f$-sum rule in Eqn.(4) above now allows one to extract $\omega_{H}$ even in 
the regime sans any Landau quasiparticles, since the existence or non-existence of the Landau 
quasiparticle concept does not enter anywhere in the evaluation of $\omega_{H}$ above. Thus, in 
the strange metal, insistence on retaining the notion of cyclotron orbits leads to a novel
conclusion: the cyclotron orbits have, as one would expect, nothing to do with any traditional 
notion of a band mass in a magnetic field. However, with a suitably re-interpreted $\omega_{H}$ 
as above, one can still work within the structure elaborated above. In the correlated LFL metal 
that obtains when the interband one-electron hybridization is relevant, one correctly recovers 
a renormalized cyclotron frequency, $\omega_{H}^{FL}=\omega_{c}^{*}=eB/m^{*}$ with large but 
finite $m^{*}$. This reinterpretation of cyclotron orbits in a local critical metal must, 
to be internally consistent, have novel implications for magnetotransport in the strange metal: 
in particular, the $\omega,T$ dependent Hall constant and angle will have nothing to do with 
the vagaries (shape and size) of the Fermi surface. There is limited 
theoretical~\cite{pwa,shastry}and clearer experimental~\cite{ong} support for this belief in the 
normal state of cuprate metals. We believe that elaboration of these features will require a 
comparably drastic modification of extant semiclassical ideas, and leave it for future work.

\vspace{0.5cm}

\noindent {\bf Disorder effects and modified Dingle scaling in the strange metal:}
Within our approach, we can also study the non-perturbative effects of chemical disorder 
and modification of text-book Dingle scaling in the strange metal. This is because within 
the structure of DMFT within which the local critical metal was obtained, the single-site 
coherent potential approximation (CPA) solves the Anderson disorder problem {\it exactly}, so 
using CPA with the branch-cut propagator as an unperturbed propagator in the strange metal 
allows clean conclusions within a well-controlled scheme to be drawn.  Specifically, 
consider $x$ impurities randomly distributed in the system, entailing the 
addition of a term $H_{dis}=\sum_{i}v_{i}n_{ic\sigma}$ to the EPAM. We choose 
a binary alloy disorder distribution $P(v_{i})=(1-x)\delta(v_{i})+x\delta(v_{i}-v)$. Quite 
generally, the Dingle factor represents the damping of the dHvA signal due to incoherent disorder
scattering.  As long as a correlated FL metal obtains, (i) the adiabatic continuity, and (ii) the 
unitarity limit for strongly correlated FL, which in DMFT implies invariance of the DOS at 
$\omega=0$, independent of interaction strength in the disorder-free limit, can be used to 
treat {\it weak} disorder effects perturbatively using the self-consistent Born 
approximation. The leading effect will then be a finite but constant scattering rate, 
$\tau^{-1}=$Im$\Sigma_{imp}(\omega)=-v^{2}x(1-x)/W$, with $W$ the disorder-free bandwidth. This 
immediately accounts for the text-book Dingle scaling when one plugs in $\tau$ above into $R_{D}(T)=$exp$(-\pi/\omega_{c}\tau)$, where $\omega_{c}=(eB/m_{b})$ with $m_{b}$ the {\it bare} 
band mass and $B$ the magnetic field. For stronger disorder in the strongly 
correlated metal with very small $z_{FL}$, one must already go beyond SCBA, and 
DMFT+CPA yields incoherent non-FL metallic states going over to Mott-Anderson 
insulating states as a function of disorder~\cite{laad-IPT+CPA}. 

Quite remarkably, as with the notion of cyclotron mass, this standard picture is 
completely modified in the strange metal. One might have expected this on qualitative 
grounds: since the strange metal has in-built quantum-critical features and a $z_{FL}=0$, {\it 
any} perturbation is expected to be relevant.  Quite simply, when $z_{FL}=0$, even a small 
bare disorder potential cannot be treated perturbatively, because the wipe-out of the FL 
coherence scale ($T_{coh}=0$) means complete invalidity of any perturbative-in-
$v/k_{B}T_{coh}$ regime. Put another way, assuming scattering only in the $s$-wave channel, 
(non-magnetic disorder) a rigorous Ward identity, Lim$_{q\rightarrow 0}\Lambda_{q}
(k_{F},\omega=0)=z_{FL}^{-1}$~\cite{kotliar} implies an intrinsically divergent vertex, 
precluding use of any perturbative approach. Thus, there is no reason to expect any 
conventional scaling anymore. To see how traditional Dingle scaling is modified, we have 
to delve deeper into the structure of CPA equations with a quantum-critical bare propagator.
Quite generally, given the $G(k,\omega)$ for a pure ``reference'' medium (in our case the 
strange metal), the CPA gives a configurationally averaged self-energy $\Sigma_{imp}(\omega)$ 
(and Green function) as 
\begin{equation}
\Sigma_{imp}(\omega)=-vx-\frac{v^{2}x(1-x)}{\omega-v(1-x)-\sum_{k}t_{k}^{2}G(k,\omega)}
\end{equation}

\noindent and $G_{imp}(\omega)=\frac{1-x}{\omega-\sum_{k}t_{k}^{2}G(k,\omega)}+\frac{x}{\omega-v-\sum_{k}t_{k}^{2}G(k,\omega)}$.  Here, $G^{-1}(k,\omega)=\omega-\epsilon_{k}-a|\omega|^{1-\alpha}-\Sigma_{imp}(\omega)$ in the strange metal (notice that
$\Sigma_{imp}(\omega)$ itself enters the right hand side of Eqn.(2) itself in a
 fully self-consistent treatment).  Noticing that Eqn.(2) can be re-written as 
$\Sigma_{imp}(\omega)=-vx-v^{2}x(1-x)G_{bb}(\omega-v(1-x))$, and that close 
to the Fermi energy,
 we can neglect both $\omega$ in the denominator of Eq.(2) along with Re$\Sigma(\omega)$ in comparison with the large 
Im$\Sigma(\omega)\simeq \omega^{1-\alpha}$,
we finally find Im$\Sigma_{imp}(\omega)\simeq -vx-\frac{v^{2}x(1-x)}{a\omega^{1-\eta}}\Psi[\frac{1}{2}-\frac{i\omega}{2\pi T}]$ with $\Psi(x)$ 
the digamma function.  And the now $T$-{\it dependent} Dingle factor reads

\begin{equation}
R_{D}(T)\simeq e^{[-\frac{v^{2}x(1-x)}{\omega_{c}T^{1-\eta}}]}
\end{equation}
so that we predict that log(log$R_{D}(T)$) vs log$(T)$ will show a linear 
behavior in strong contrast to the textbook form.  For a local quantum critical metal with static disorder, the 
ln(ln$\chi_{osc})\simeq$ ln(ln$T$)-$(1-\alpha)$ln$T$ behavior in the pure case will thus acquire an additional $-ln(T)$ contribution; i.e,
we predict that the $T$-dependent dHvA susceptibility will exhibit an $x$-dependent (constant) shift on the ln(ln$\chi_{osc})-$ln$T$ plot.
 This is a counter-intuitive 
result at first sight, since $R_{D}(T)$ decreases (dHvA signal gets more damped)
 as $T$ is reduced, in contrast to the influence of the thermal factor implied
in Eqn.(1), which leads to increased damping as T is raised.  Upon closer 
reflection, we however see that this dependence is just a manifestation of the 
fact that, as in $D=1$  Luttinger liquids, the effect of disorder in the strange metal is even more singular as argued above,  and drives 
a direct transition to a ``strange insulator".  In fact, such an eventuality was foreshadowed in earlier work, 
both in the tomographic Luttinger liquid~\cite{pwa-rama} and marginal-FL~\cite{cmv-1997} contexts.  Our work is a rigorous theoretical 
scenario for the high-$D$ local critical metal, and our specific prediction for this effect in dHvA oscillations is that the Dingle factor 
will enter as above, i.e, in a way drastically different from the text-book form.  The new -ln$T$ correction due to disorder could be 
unearthed by intentional non-magnetic chemical substitution in CeCoIn$_{5}$~\cite{julian} 
or YbRh$_{2}$Si$_{2}$~\cite{steglich} if quasilocal criticality indeed 
underlies their novel responses.  Our findings could also
conceivably be relevant to study of impurity effects in the context of dHvA 
studies for strange metals in a broader context, and, in particular, should 
aid in interpretation of FS changes as a strongly correlated FL metal is driven through a ``local'' quantum critical point(phase) by suitable tuning.

  The above provides an explicit prediction of the dHvA effect in the ``local 
quantum critical'' phase that is generically expected to appear close to the 
selective-Mott quantum phase transitions. The central underlying reasons for the
 radical departure from traditional behavior within the structure of the DMFT 
used here is the
Kondo destroying extinction of the FL coherence scale in the OSM phase.  This 
necessitates abandonment of any ``analytic continuation'' of semiclassical 
transport formalism(s) from the outset, all of which rely primarily on a 
suitably renormalized but finite band mass. 
\vspace{0.3cm} 

\noindent {\bf Angle Resolved Photoemission}: 
As emphasized in the introduction, it is important that internal consistency between dHvA and ARPES data be 
achieved for a comprehensive understanding of fermiology in non-FL metals. Let us now consider how complementary 
ARPES spectra reflect changes across an OSMT. The structure of the $a,b$ band propagators in two-orbital DMFT 
is such that $G_{\alpha}^{-1}(k,\omega)=[\omega-\epsilon_{\alpha}(k)-\Sigma_{\alpha}(\omega)-
\frac{(V_{\alpha\beta}(k)-\Sigma_{\alpha\beta}(\omega))^{2}}{\omega-\epsilon_{\beta}(k)-\Sigma_{\beta}(\omega)}]$.
In the heavy-FL metal for $U_{fc}<U_{fc}^{(1)}$~\cite{laad1}, both $G_{a},G_{b}$ have the (same) pole structure, 
determined from the equation

\begin{equation}
\omega-\epsilon_{\alpha}(k)-\Sigma_{\alpha}(\omega)=\frac{(V_{\alpha\beta}(k)-\Sigma_{\alpha\beta}(\omega))^{2}}{\omega-\epsilon_{\beta}(k)-\Sigma_{\beta}(\omega)}
\end{equation}

  However, the spectral fraction (of $a$-or $b$) contribution to the bands is quite involved, reflecting 
the redistribution of weight in each band, caused by non-trivial frequency-dependent structure of the 
off-diagonal self-energy, $\Sigma_{ab}(\omega)$. In the HFL metal, the ``$a$'' band which crosses 
$E_{F}$ has predominantly $f$ character and a heavy mass.  As shown in Fig.~\ref{fig2} a,b, the band structure 
in this limit is a heavily renormalised mirror of the two hybridised bands of the EPAM. The severely narrowed 
coherent part of the $b$-band, clearly seen as a dispersive feature around the K point with a small band 
width, attests to the heavy FL state with small $z_{FL}$.  At higher energies, the Hubbard ``bands'' are visible 
as weakly dispersive and strongly damped features carrying a large fraction of the total spectral weight. 
Interestingly, a correlation-induced low energy kink is also clearly visible in Fig.~\ref{fig2}a: its energy scale 
$\omega_{\alpha}^{*}$ is interpreted as that corresponding to the crossover between low-energy coherent and 
higher energy incoherent features in the spectral function of a strongly correlated FL, in accord with 
earlier work. A small $\omega^{*}$ is characteristic of drastically reduced $z_{FL}$ and low 
FL coherence scale, in accord with what is expected in a heavy FL metal.

What happens across the OSMT?  Since Im$G_{aa}(k,\omega)$ and Im$\Sigma_{a}(\omega)$ now show Mott insulating 
features (the latter exhibiting a pole at the Fermi energy), the $a$ fermions do not contribute anymore to the 
Luttinger volume of the HFL phase.  Now, the low-energy structure is determined from the solution of the equation

\begin{equation}
\omega^{1-\alpha}-\epsilon_{b}(k)=\frac{(t_{ab}(k)-\Sigma_{ab}(\omega=\mu))^{2}}{\omega-\epsilon_{a}(k)-U_{ab}^{2}/(\omega+\mu)}
\end{equation}

\begin{figure*}[h]
(i)
{\includegraphics[angle=0,width=0.5\columnwidth]{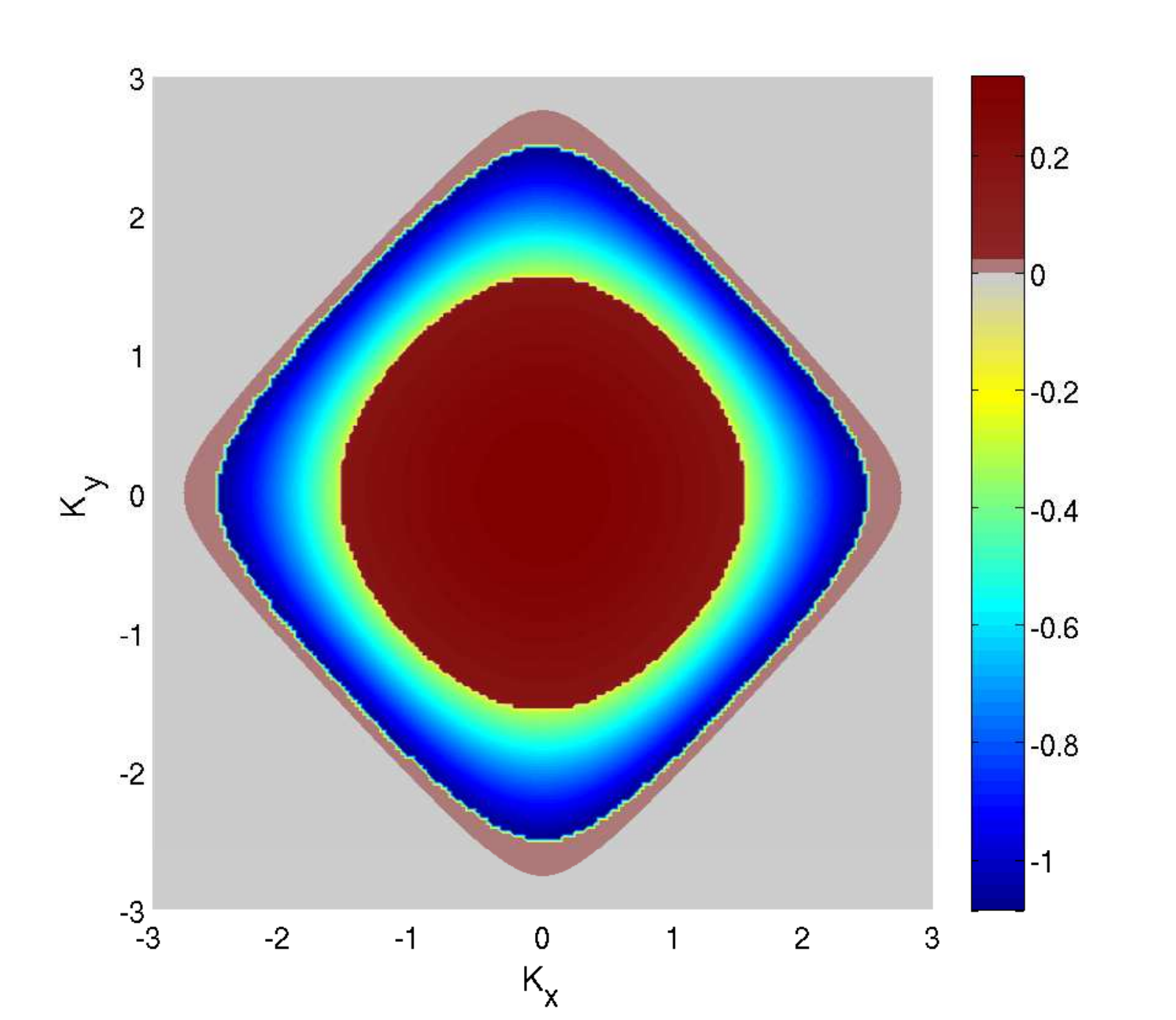}}
(ii)
{\includegraphics[angle=0,width=0.5\columnwidth]{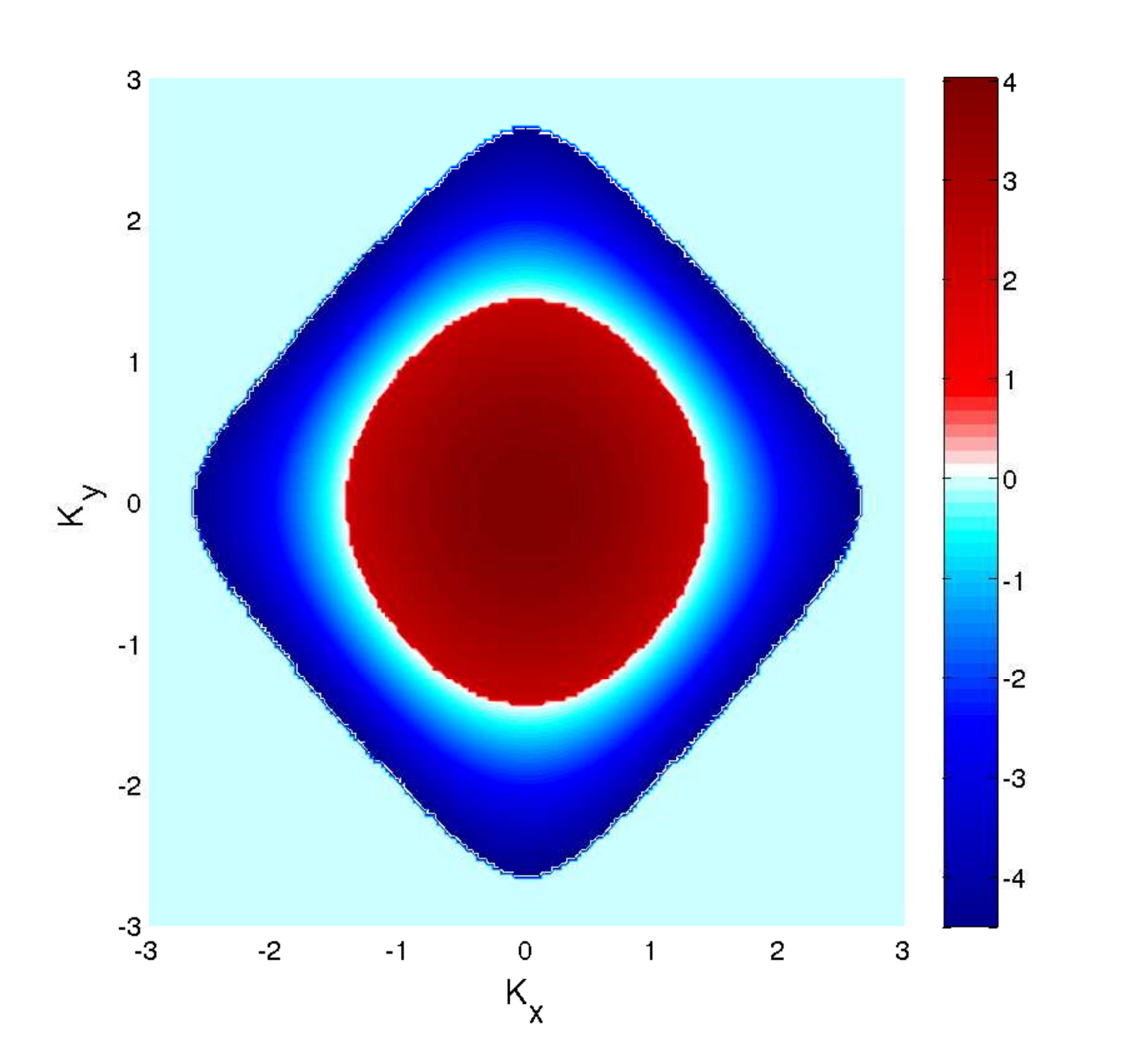}}
(iii)
{\includegraphics[angle=0,width=0.5\columnwidth]{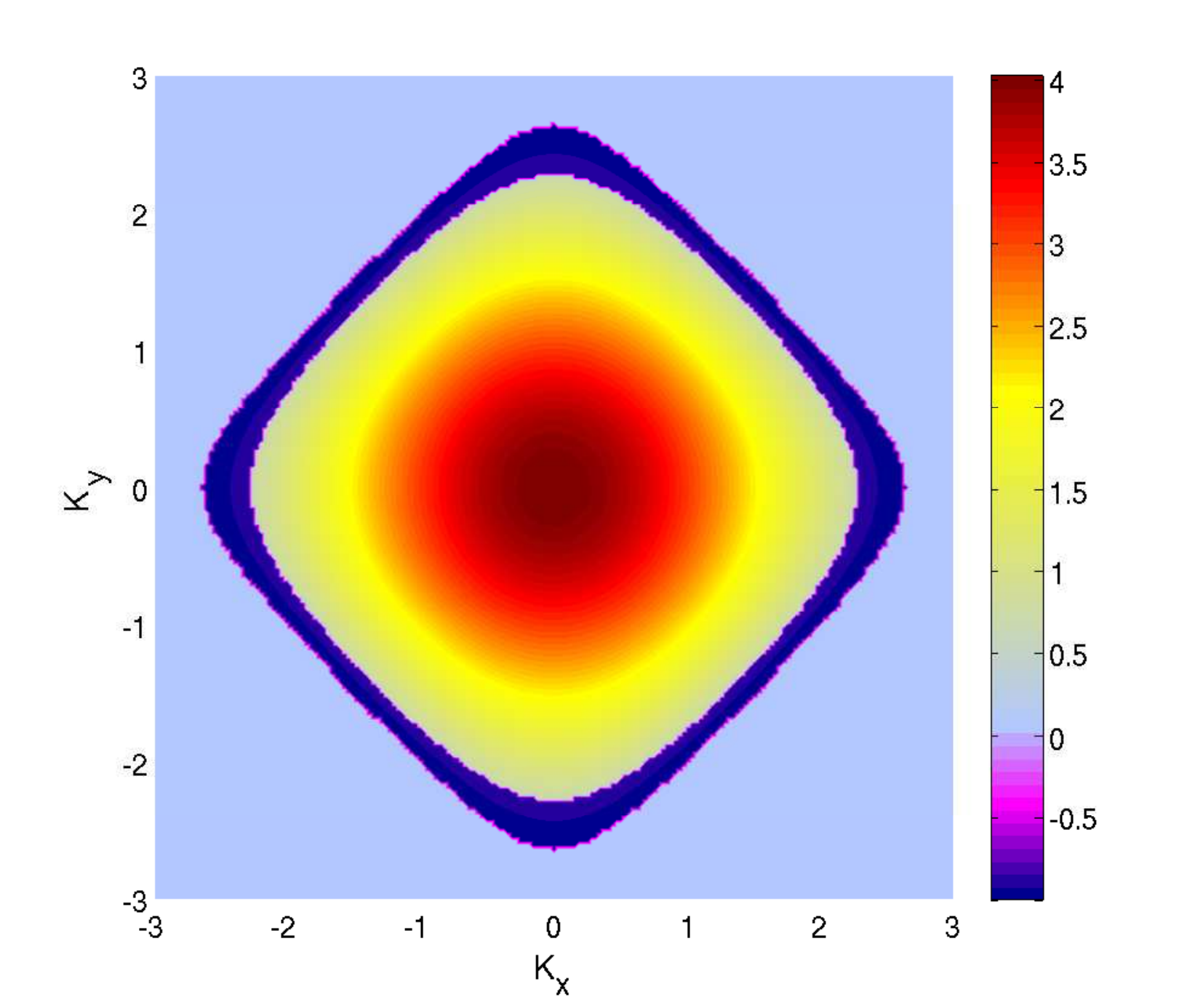}}

(iv)
{\includegraphics[angle=0,width=0.5\columnwidth]{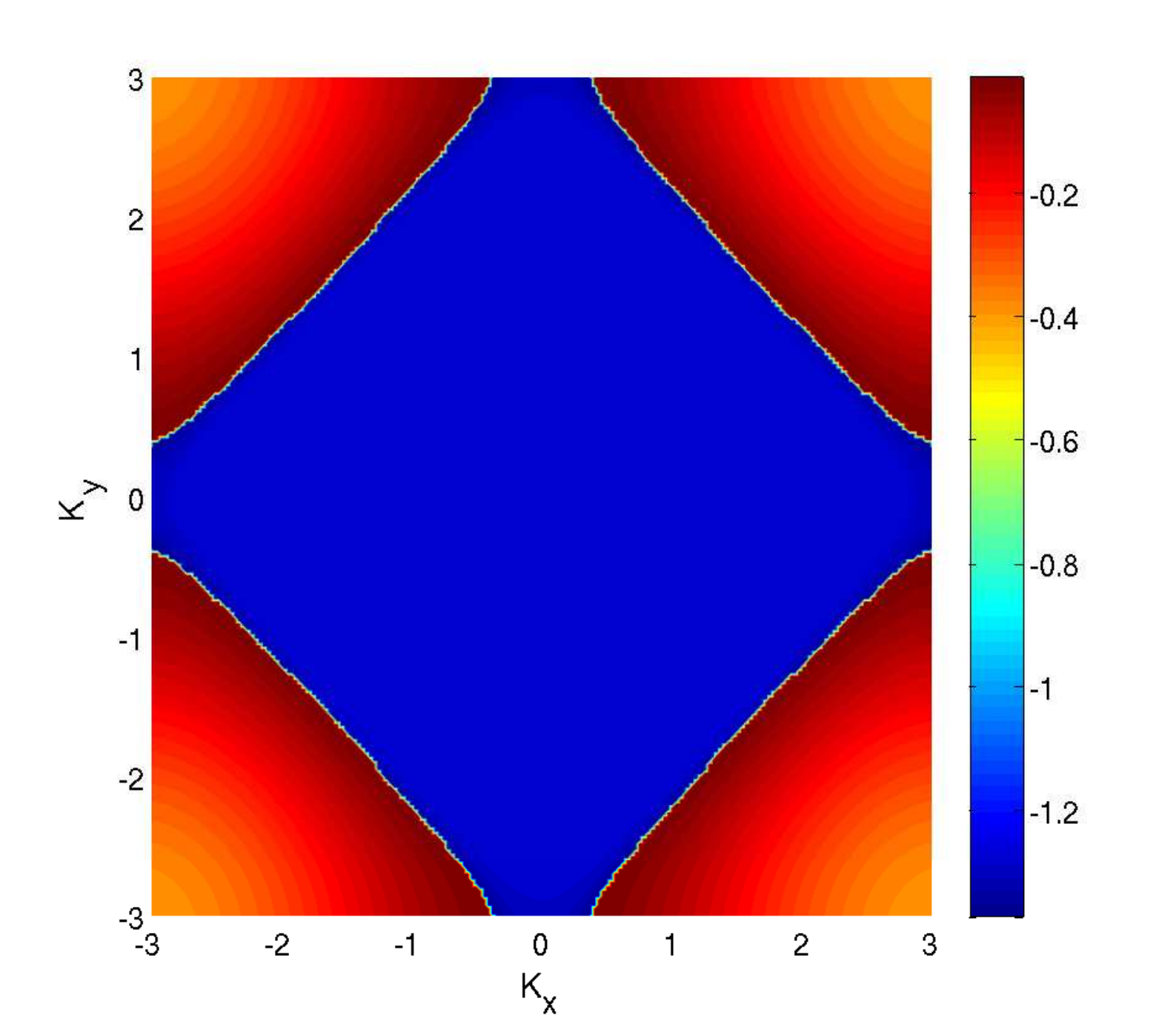}}
(v)
{\includegraphics[angle=0,width=0.5\columnwidth]{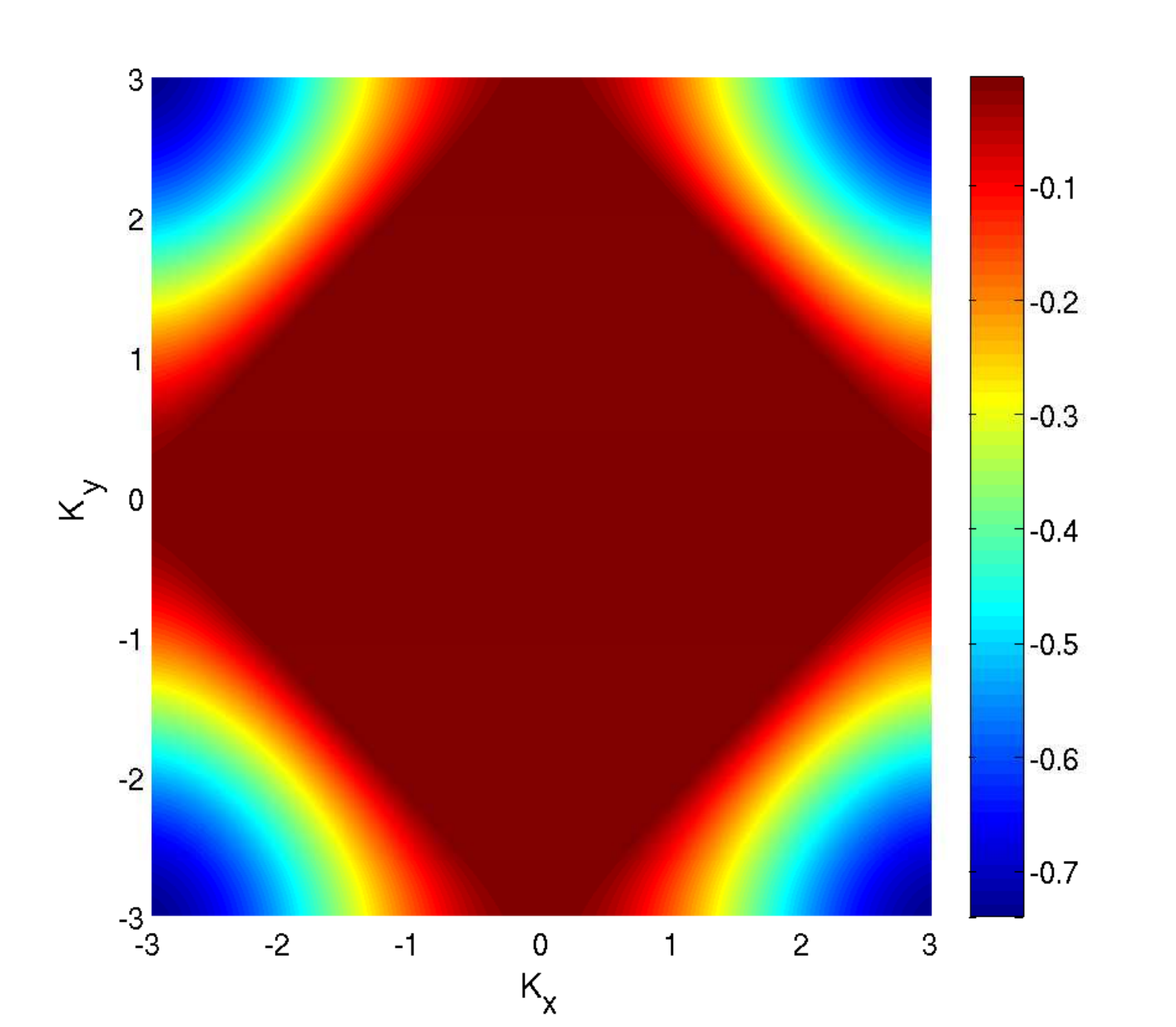}}
(vi)
{\includegraphics[angle=0,width=0.5\columnwidth]{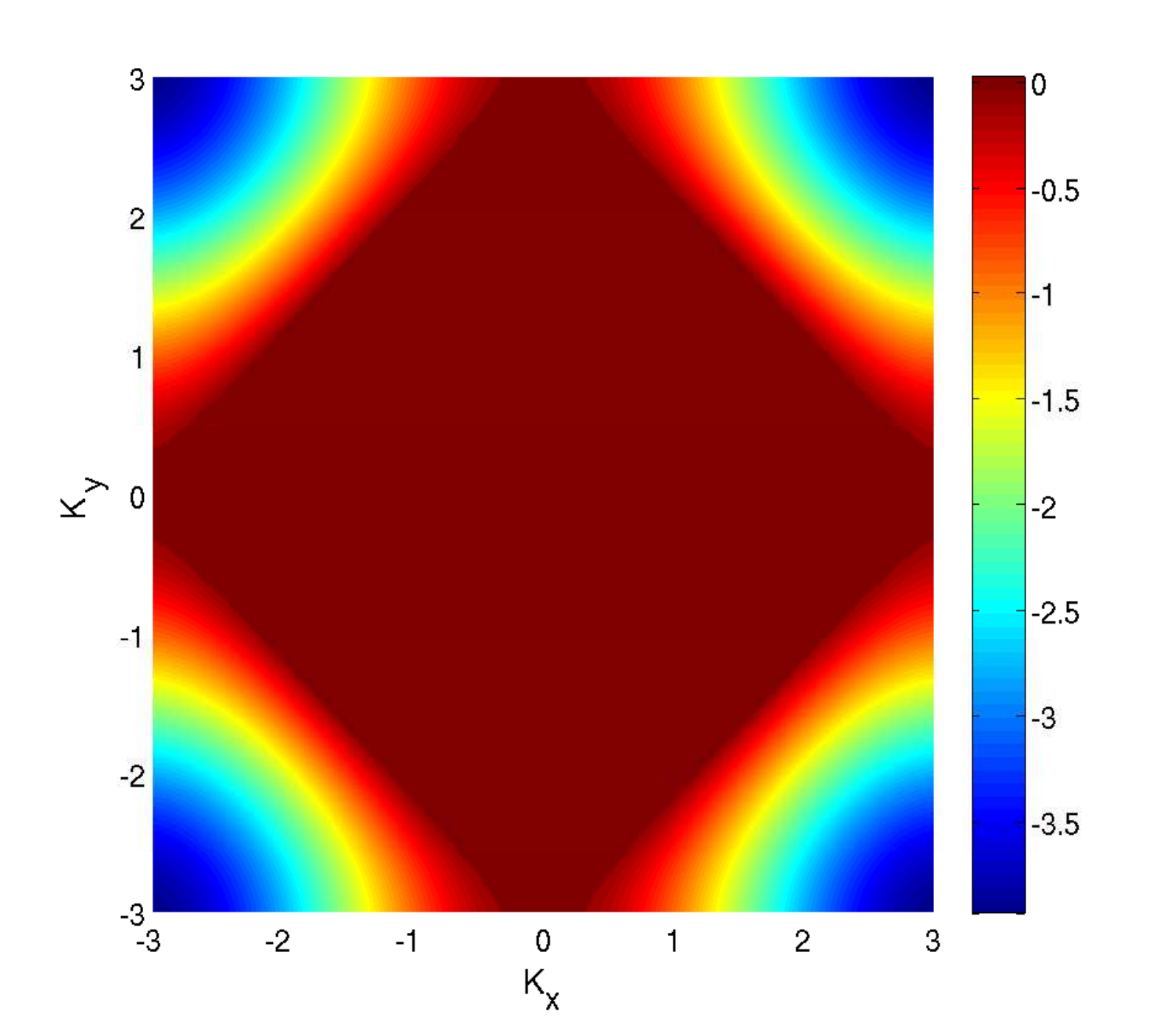}}
\caption{(Color Online)FS plots for (i),(iv)U=3, U’=0, (ii),(v)U=3, U’=0.5, (iii),(vi)U=3, U’=1.5. Upper pannel is for the $a$ band and lower panel for the $b$ band.  In the heavy-Fermi liquid phase [(i)-(iv) and (ii)-(v)], both $a,b$-band Fermi surfaces are well defined and obey the traditional Luttinger theorem.  In the OSMT phase,
however, the $a$-``Fermi surface'' now corresponds to the {\it zeros} of $G_{aa}(k,\omega)$, while the $b$-Fermi surface is critical in the sense that it is still formally well-defined (since Im$\Sigma_{bb}(\omega=0)=0$) even when Landau Fermi liquid quasiparticles are absent (since $z_{FL}=0$, see text for details).}
\label{fig1}
\end{figure*}

\begin{figure*}[h]
(i)
{\includegraphics[angle=0,width=0.5\columnwidth]{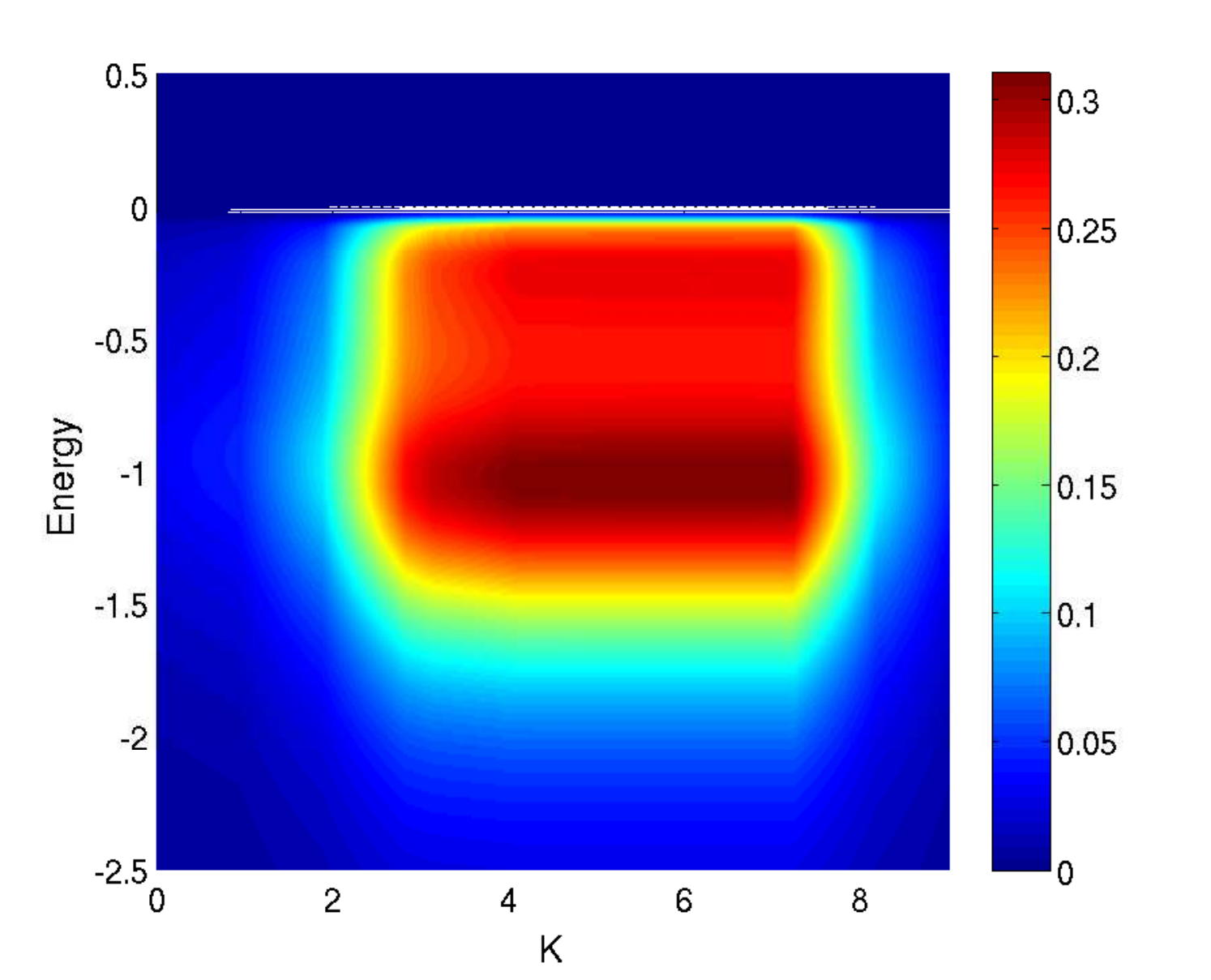}}
(ii)
{\includegraphics[angle=0,width=0.5\columnwidth]{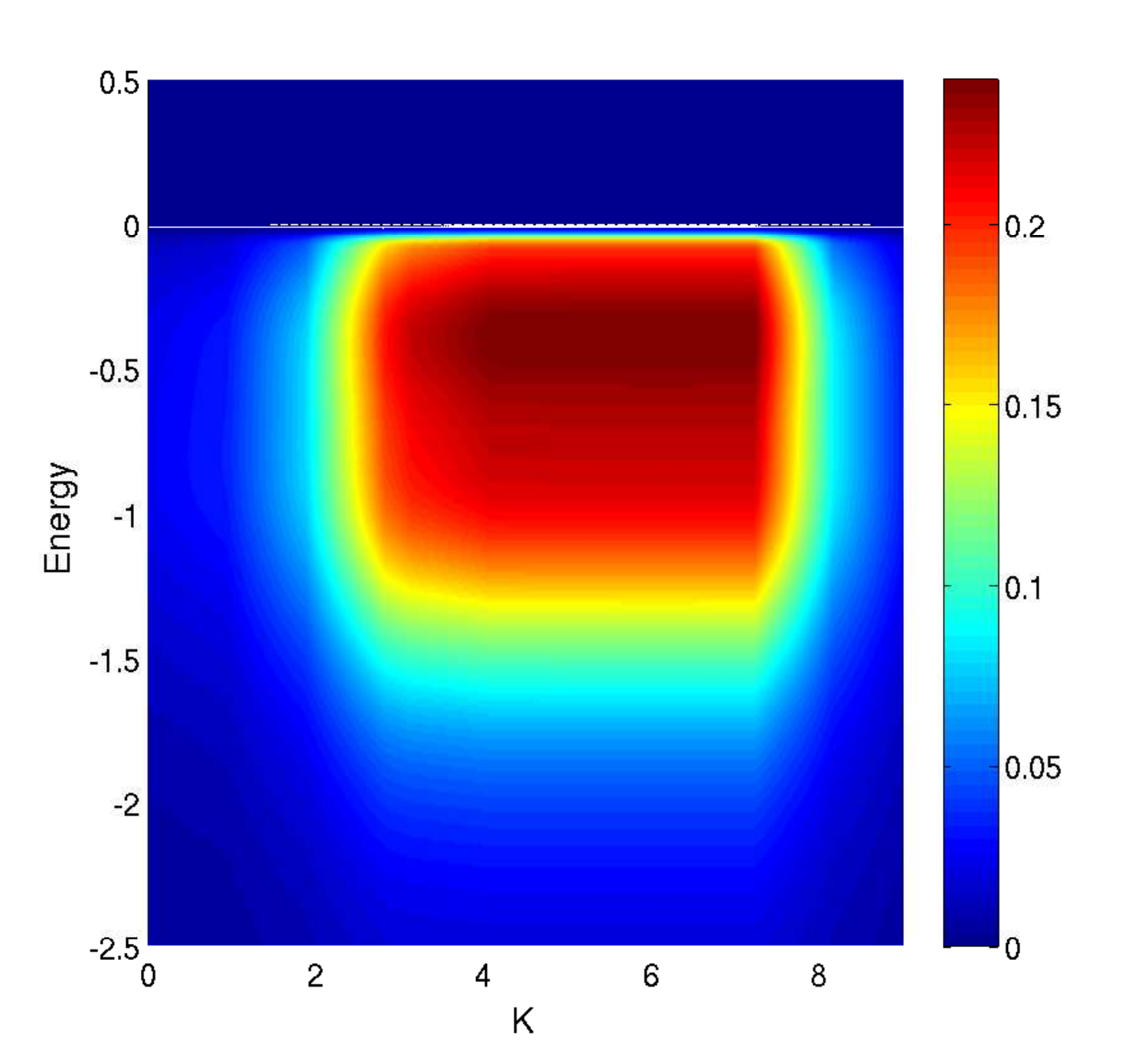}}
(iii)
{\includegraphics[angle=0,width=0.5\columnwidth]{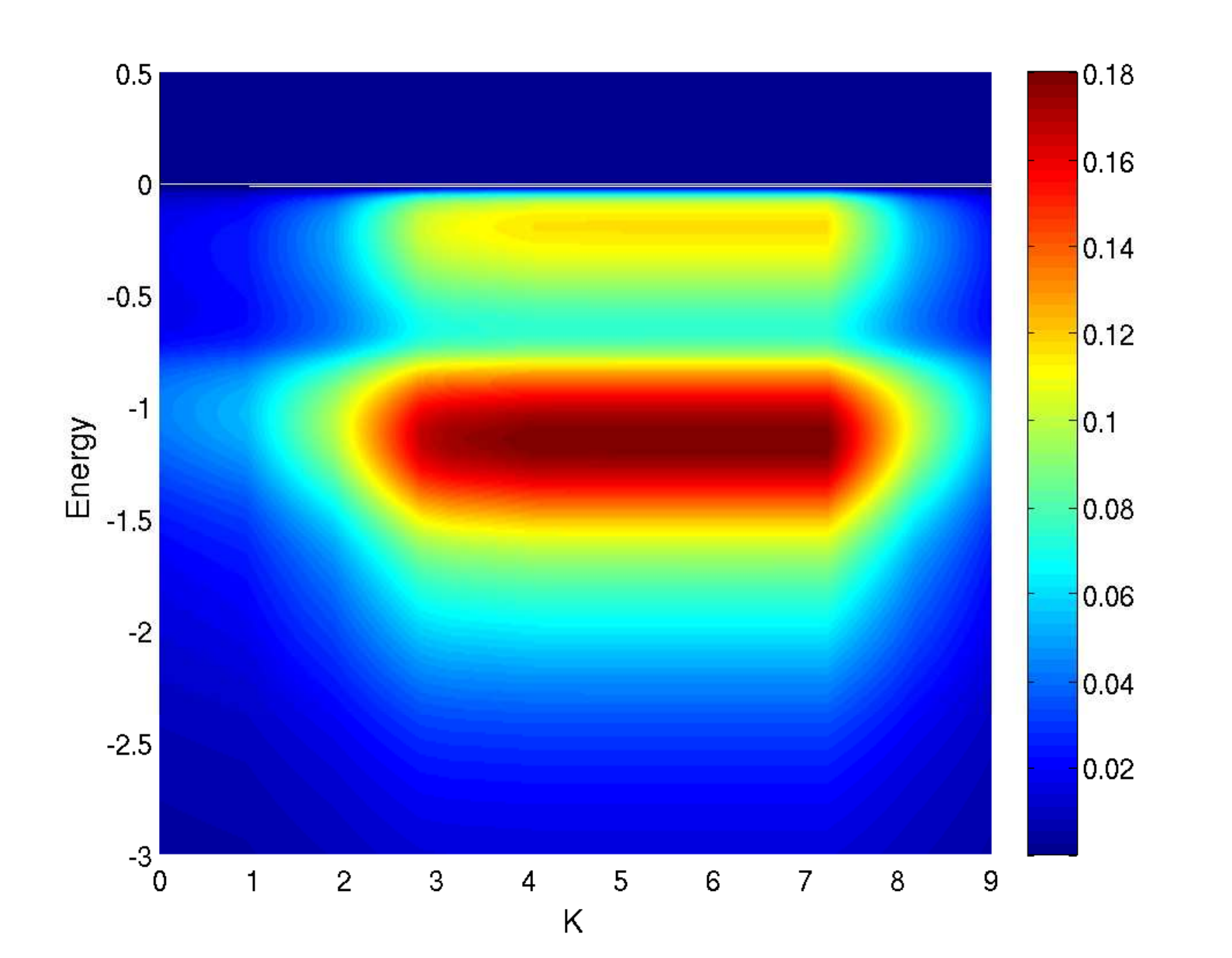}}

(iv)
{\includegraphics[angle=0,width=0.5\columnwidth]{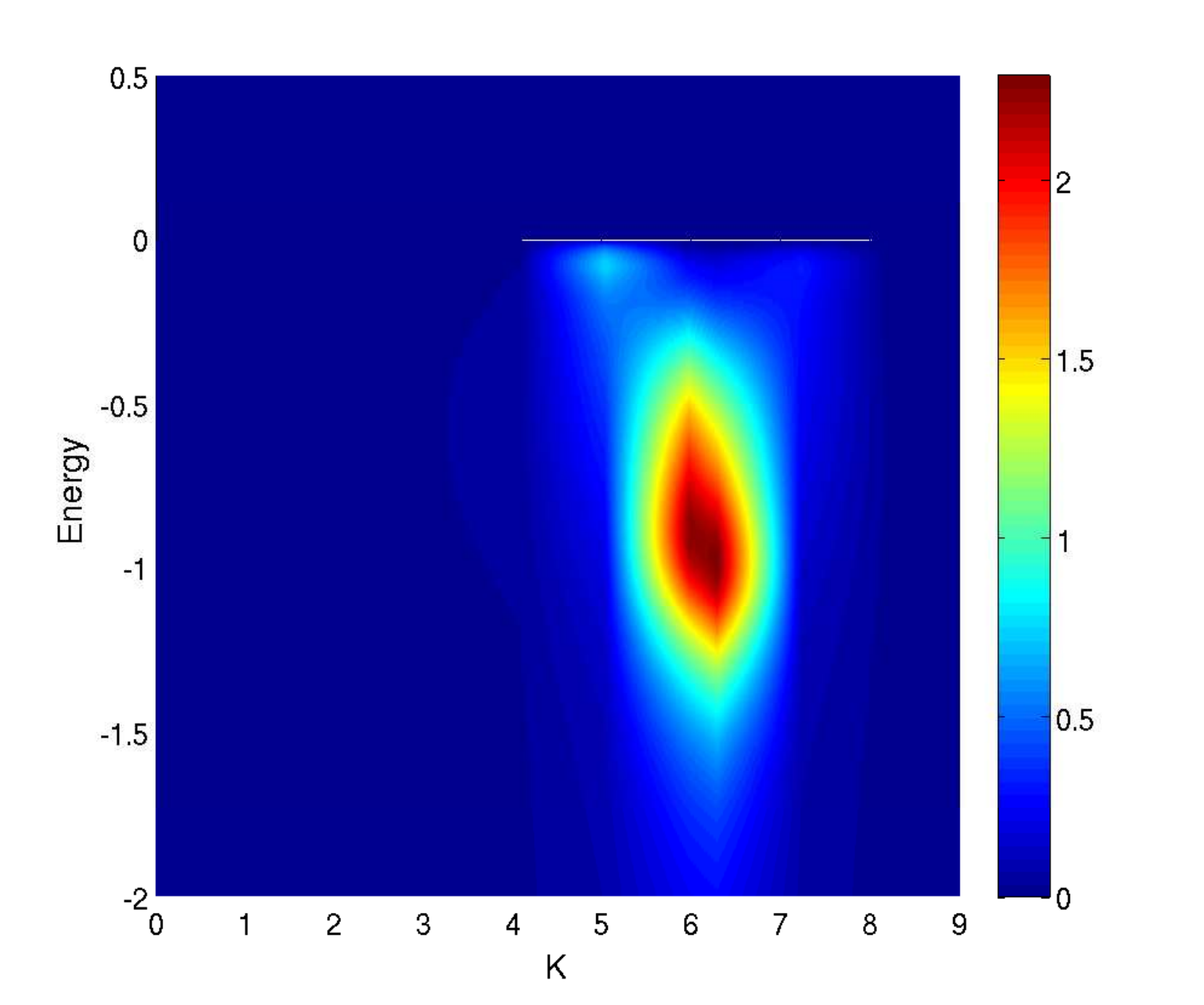}}
(v)
{\includegraphics[angle=0,width=0.5\columnwidth]{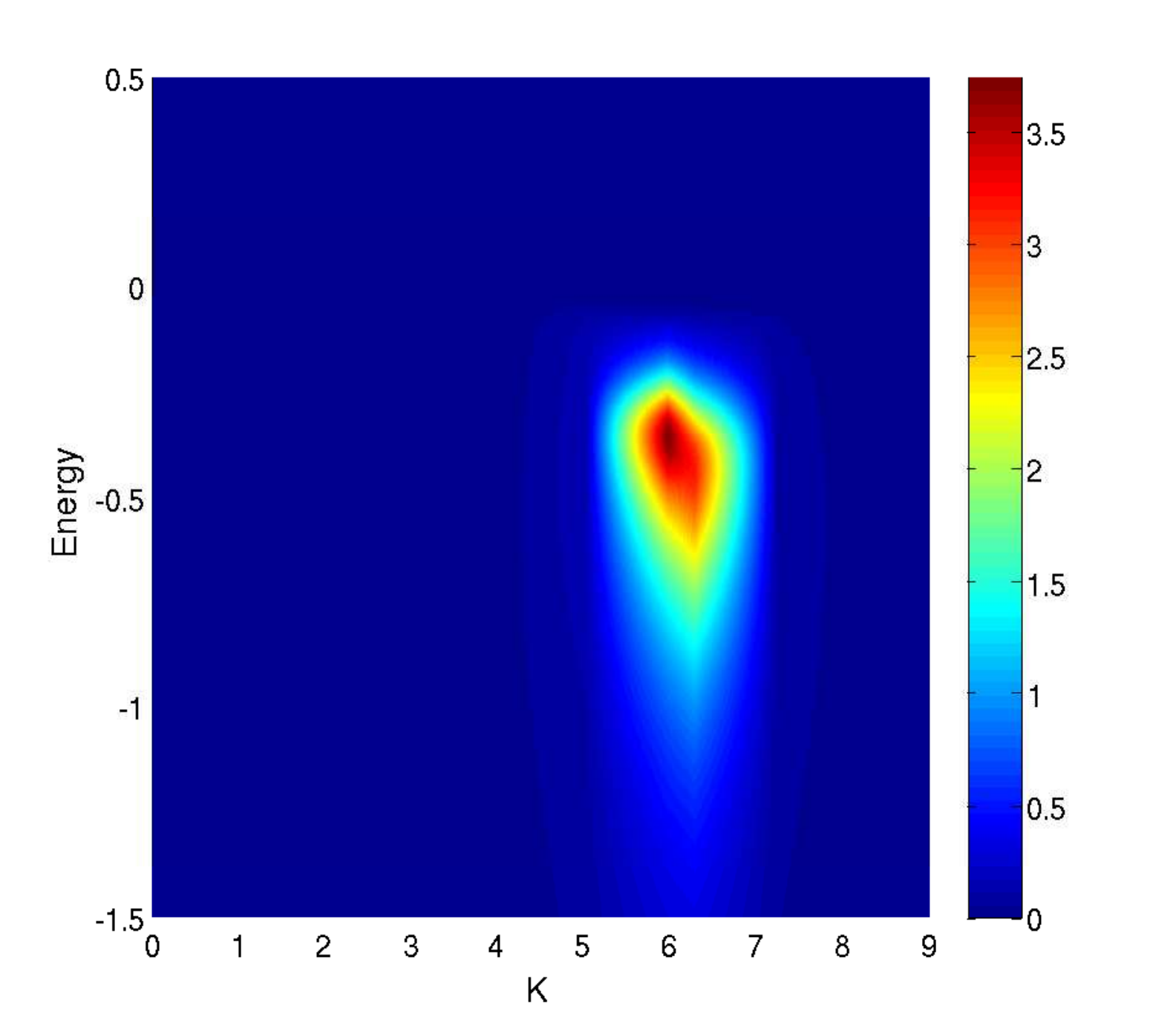}}
(vi)
{\includegraphics[angle=0,width=0.5\columnwidth]{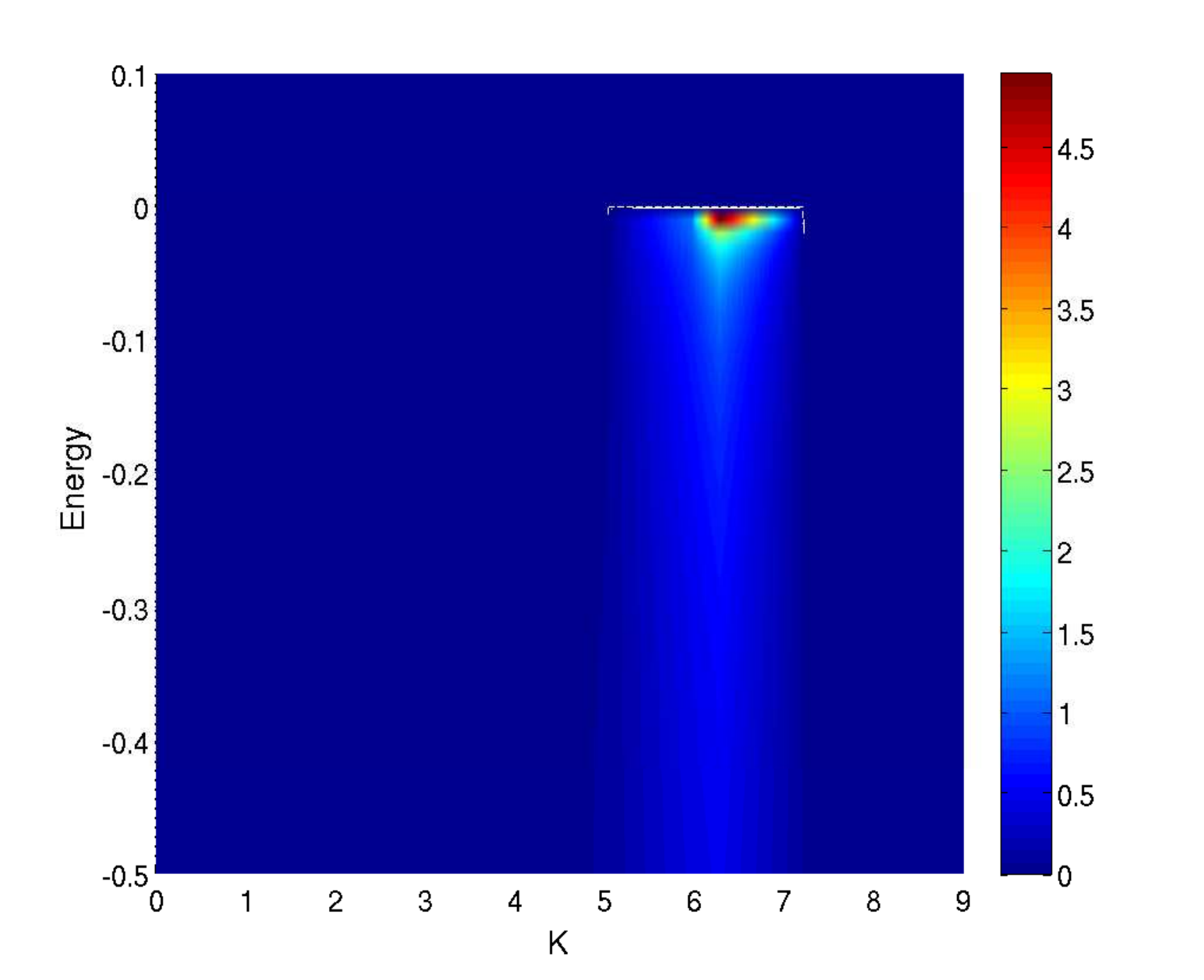}}
\caption{(Color Online)ARPES plots for (i),(iv)U=3, U'=0, (ii),(v)U=3, U'=0.5, 
(iii),(vi)U=3, U'=1.5. Upper pannel is for the $a$ band and lower panel for the $b$ band.  In [(i),(iv)], a clearly dispersive $b$-band with drastically reducced band width (small but finite $z_{FL}$) obtains.  In [(ii),(v)], close to the OSMT, the $b$-band dispersion is almost absent, but a small dispersion is still visible.  In
[(iii),(vi)], however, clear Mott insulating behavior for $a$-band and destruction of the $k^{2}$-term in the 
$b$-band dispersion as a result of divergent effective mass ($m_{b}^{*}=z_{FL}^{-1}$) is clearly seen (see text for details).}
\label{fig2}
\end{figure*} 

\begin{figure*}
(i)
{\includegraphics[angle=0,width=0.5\columnwidth]{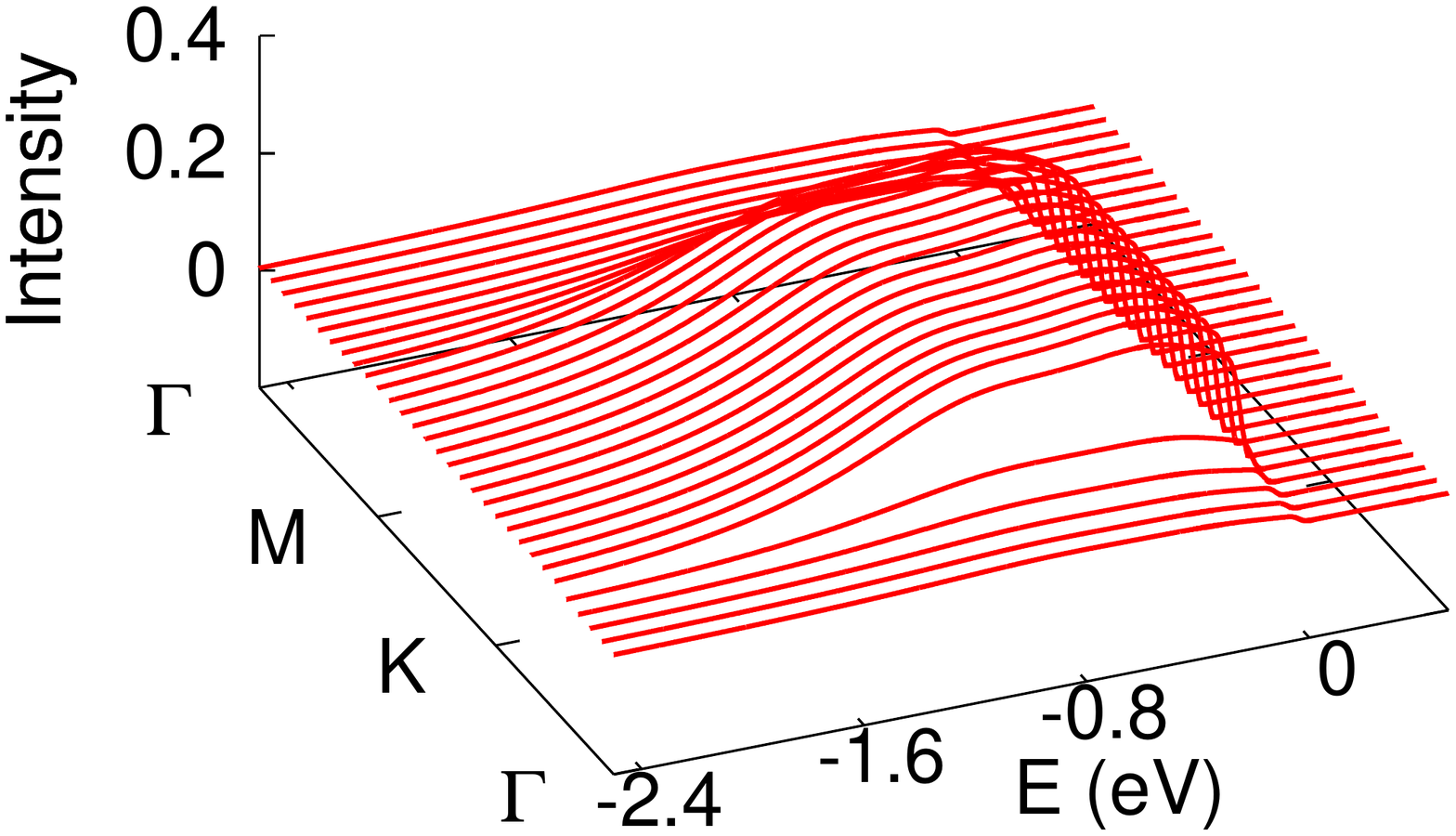}}
(ii)
{\includegraphics[angle=0,width=0.5\columnwidth]{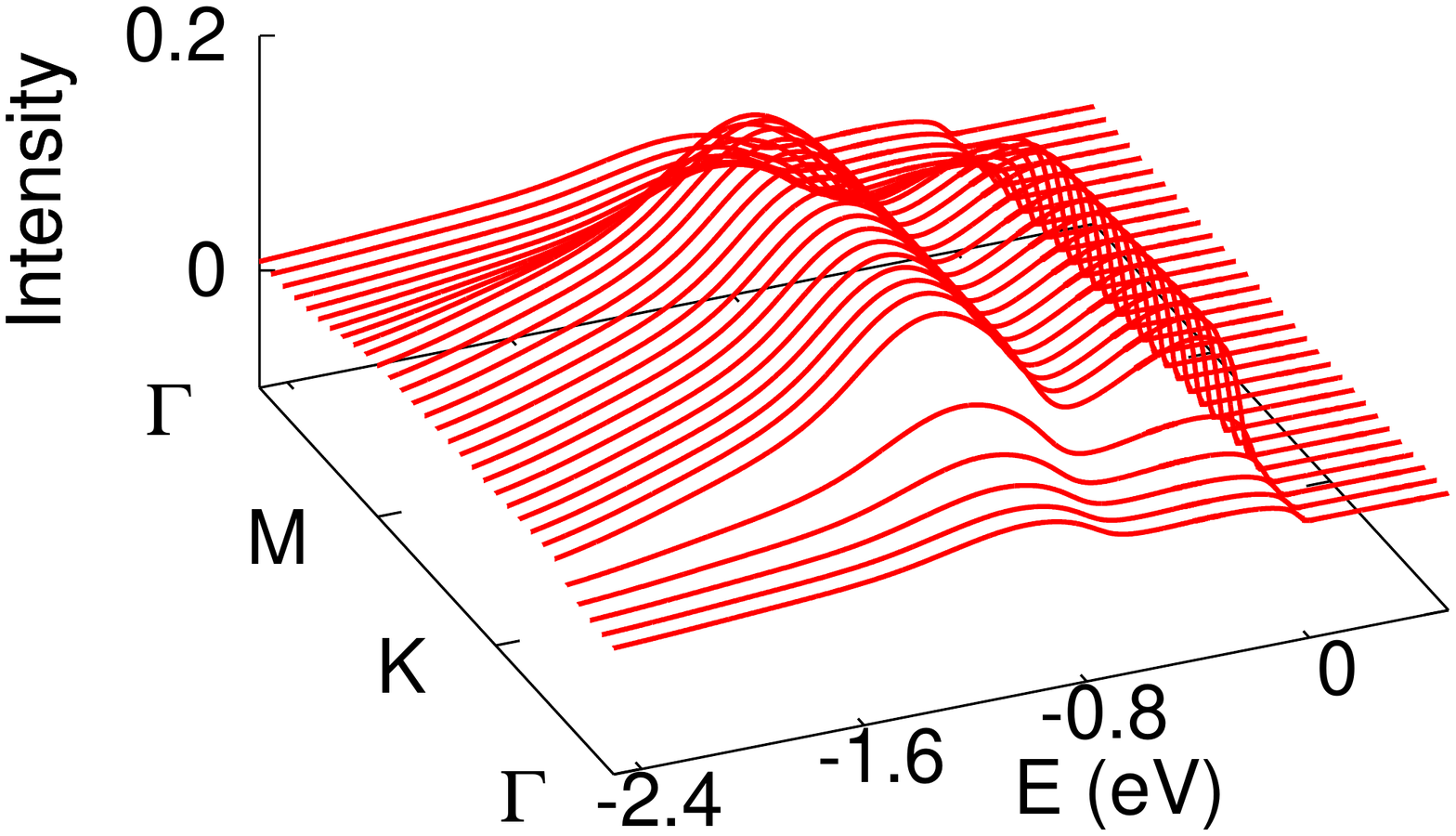}}
(iii)
{\includegraphics[angle=0,width=0.5\columnwidth]{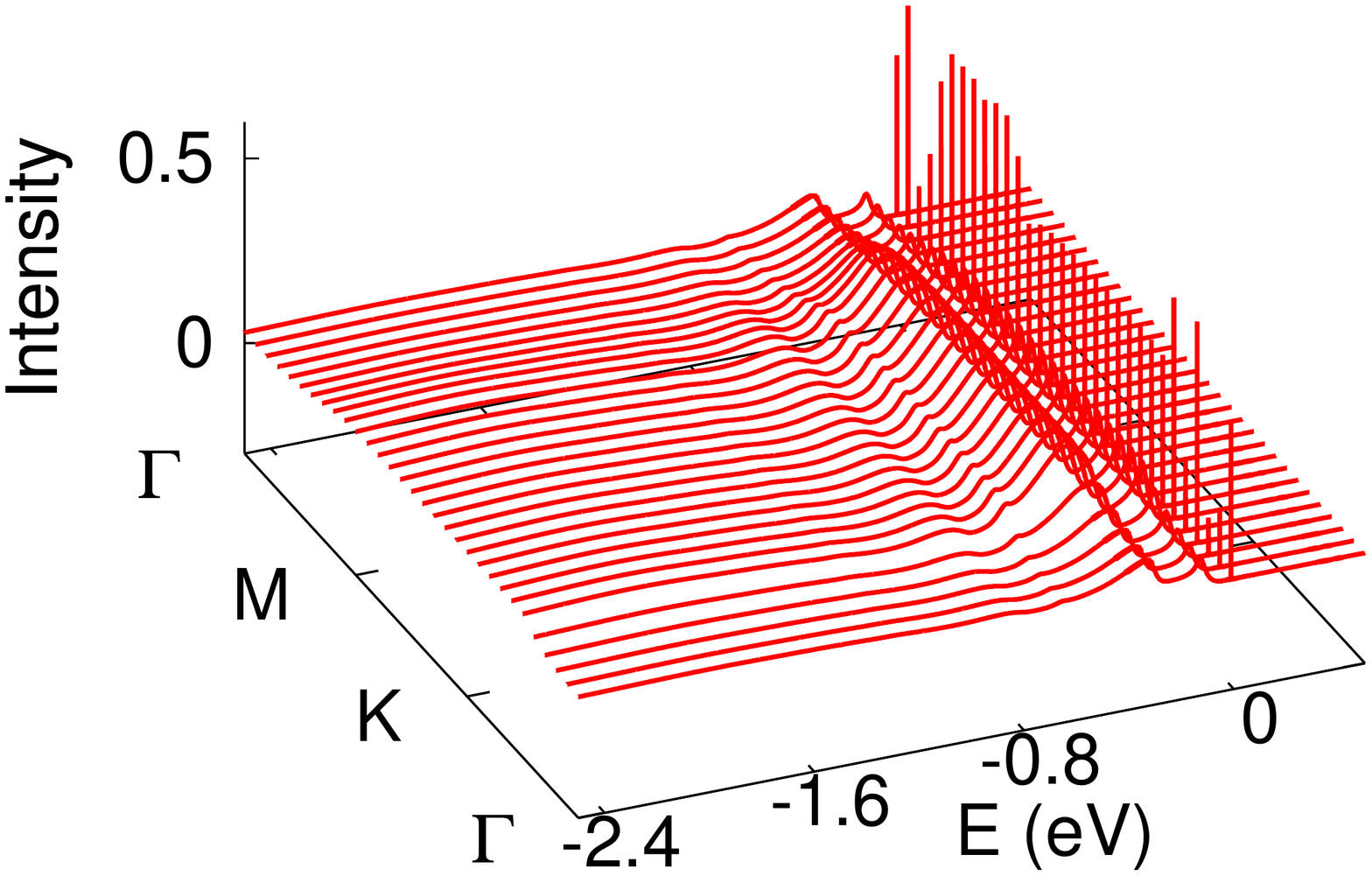}}

(iv)
{\includegraphics[angle=0,width=0.5\columnwidth]{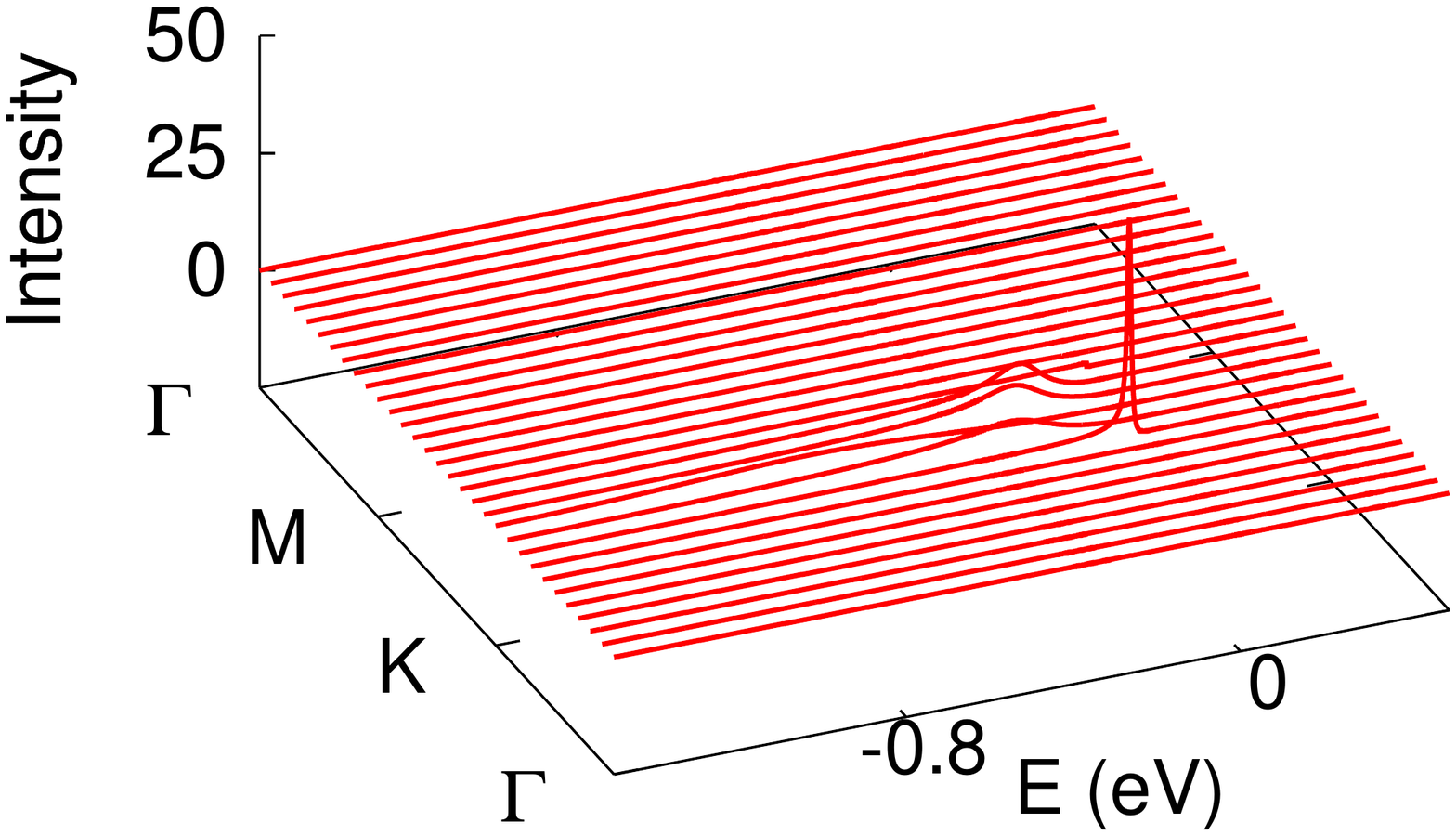}}
(v)
{\includegraphics[angle=0,width=0.5\columnwidth]{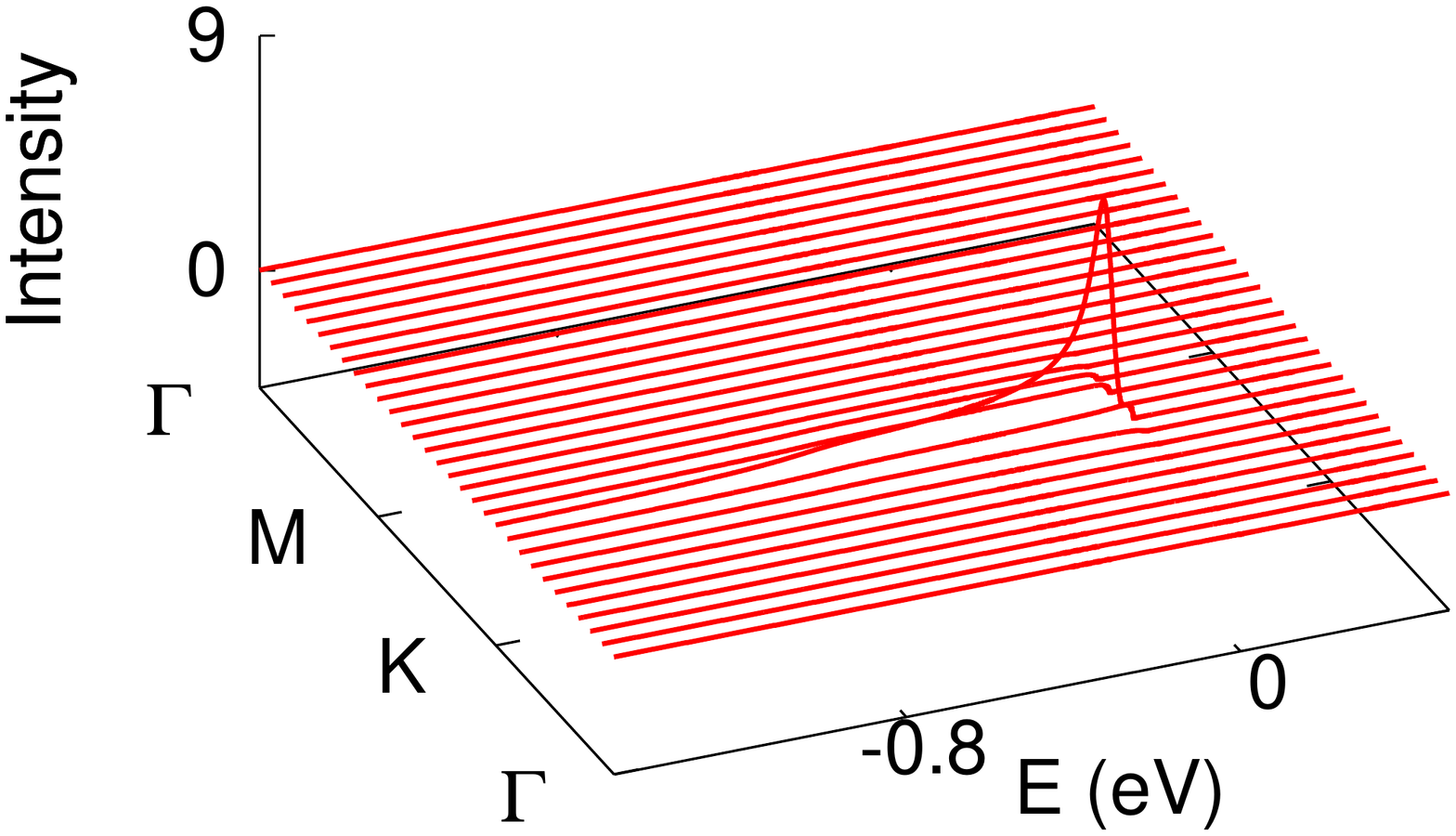}}
(vi)
{\includegraphics[angle=0,width=0.5\columnwidth]{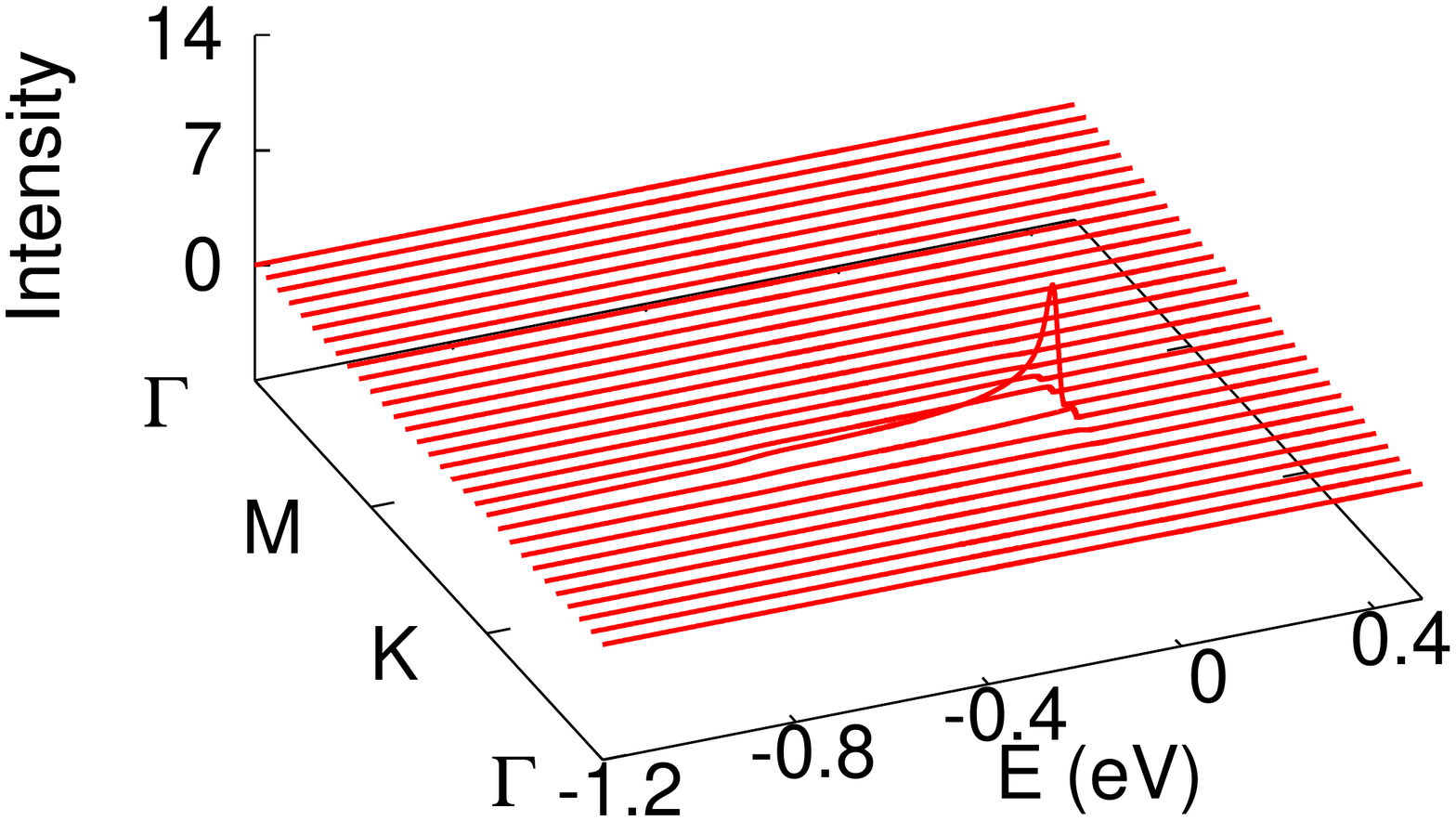}}
\caption{LDA+DMFT bands for (i,iv)U=3, U’=0, (ii,v)U=3, U’=0.5, (iii,vi)U=3, U’=1.5.  ARPES lineshapes corresponding to the heavy-Fermi liquid with a heavily renormalized but clearly dispersive $b$-band (i,iv), in proximity to the OSMT but still on the heavy-Fermi liquid side with a very small but finite $b$-band dispersion (ii,v), and in the OSMT phase (iii,vi).  In this phase, clear power-law singular feature in the $b$-band spectral function (vi) along with Mott-insulating features in the $a$-band spectrum, are clearly visible. }
\label{fig3}
\end{figure*}
 
\noindent The low-energy pole in Im$\Sigma_{a}(\omega)\simeq U_{ab}^{2}/(\omega+\mu)$ breaks Luttinger's 
theorem as a direct consequence of the OSMT.  This observation will also have novel implications for 
the dHvA signals, and whether dHvA measurements can thus show up the ``spinon'' Fermi surface (surface of 
{\it zeros} of $G_{aa}(k,\omega)$) is an intriguing aspect to ponder. In DMFT, obliteration of the heavy 
FL metal via a lattice ``orthogonality catastrophe'' {\it a la} hidden-FL theory~\cite{pwa} obtains: 
however, since Im$\Sigma_{b}(\omega)$ still vanishes, albeit anomalously (like $\omega^{1-\alpha}$), close 
to the Fermi energy, the $b$-band must still appear as a band with anomalous dispersion (since the 
effective mass diverges as $m^{*}/m=z_{b}^{-1}(\omega)$) and damping. In Fig.(2), we show the $a,b$ 
band dispersions in the local critical metal. The lower ($a$) band shows clear signatures expected of a Mott 
insulator: it is dispersionless and pushed below $E_{F}(=0)$. Surprisingly, in addition to the intense 
concentration of incoherent weight in the ``lower Hubbard band'' ($\simeq -3.0$~eV), we also find clear low-energy 
features reminiscent of the renormalised $f$-like band in the range $-1.0>\omega>0$ below the selective-Mott 
gap, along with clear low-energy kink-like structures (at $\simeq 0.2-0.3$~eV) and steeply dispersive ``waterfalls" connecting the low- and high energy parts of the spectrum. The metallic component ($b$-band) clearly exhibits 
a very anomalous form: first, it is impossible to discern any remnant of the b-band dispersion of the
(heavy) FL metal. Only intense and completely incoherent spectral weight is visible in the vicinity of the K
point in the first Brillouin Zone. The corresponding DMFT local spectral function shows an anomalously 
broadened peak with a power-law fall-off instead of a sharp FL quasi-particle pole. i
Correspondingly, Re$\Sigma_{b}(\omega)$ has infinite slope at $E_{F}$, implying $z_{FL}=0$ and diverging 
effective mass. The destruction of the dispersive features is thus a consequence of the fact that, in 
the selective-metal, the quasiparticle disperion qualitatively changes its form: it is no longer given by the 
common poles structure of Eq.(4), but is a solution of Eq.(4) obtained now by inserting the pole 
of Re$\Sigma_{a}(\omega)$ and  the branch-cut form of $G_{bb}(k,\omega)$ in the infra-red. It should now be clear 
that there is no one-to-one correspondence with any FL anymore. Put another way, this non-perturbative change is 
a consequence of the selective-Mott transition of a-fermions: when a-fermions get Mott localized, we find that 
the effective mass ($m_{b}$) diverges and so the ($k^{2}/2m_{b}^{*})$ term is wiped out from the dressed dispersion
of $b$-fermions. This explains the enhanced flatness of the b-band dispersion in k-space that we observe in 
numerics in Figs.\ref{fig2} and \ref{fig3}. This is also seen very clearly in the ARPES lineshapes along 
$\Gamma-K-M-\Gamma$ in the first Brillouin zone: a weakly dispersive feature with a drastically reduced width
$O$(0.2) eV, reflective of the severely renormalised FL resonance with tiny weight of the HFL, is seen for small
$U_{fc}$. In the strange metal for $U_{fc}=1.6$~eV, however, the lack of any band dispersion for the b-band, 
along with a power-law tail extending upto high energies $O(1.0)$~eV, are clear.

The kink and waterfall structures in the a-band dipersion are also fingerprints of correlation effects, though,
given $T_{coh}=0$ in the strange metal, they cannot anymore be associated with any coherence-incoherence crossover
as for the heavy-FL metal. It is still possible, however, to interpret these features in terms of microscopic pro-
cesses underlying detailed features seen in DMFT spectra and self-energies: close examination shows distinct
structures in Im$\Sigma_{a}(\omega)$ precisely around $\omega^{*}\simeq 0.2$~eV, where the low-energy kink and 
waterfall structures obtain. From the structure of the diagrams contributing to the DMFT self-energies, we 
infer that these structures arise from the coupling of fermions to collective, coupled spin-and-charge excitations 
arising from the second-order bubble diagrams in $U_{ff},U_{cc}$, and, more importantly, $U_{fc}$
In fact, it is the last term that leads to the inverse orthogonality catastrophe resulting in $G_{bb}(\omega)$ 
above when $V_{fc}$ is irrelevant. Since the charge gap exists in the $a$-fermion sector in the OSMT phase, the 
infra-red singular branch-cut behavior as well as the kink structures in the strange metal must involve
coupling to the soft local spin fluctuations arising from the wildly fluctuating $a$-local moment of 
the (selective) Mott-Hubbard paramagnet (we have assumed no AF order to begin with). The waterfall structures, 
on the other hand, are interpretable as rapid crossovers in $k$-space between the low- and high-energy features 
(well-separated near the Mott transition in the a-fermion sector) in the DMFT spectral functions. Since 
the self-energies are $k$-independent, the momentum-space location of these features are close to the $k$-points 
where the a, b bands have zeros in the free band structures (the ones induced by Im$\Sigma_{a}$ are additional, 
correlation-induced zeros in $G_{aa}(k,\omega)$).  In DMFT, therefore, these features are intimately tied down 
to the form-factor of the inter-band hybridization, assumed here to have a $d$-wave form.

What about the Fermi surface itself? In the HFL phase, analytic continuation to a free fermion system and
the local nature of self-energies imply that the Luttinger theorem in its local version is obeyed, i.e, that the 
Fermi surface retains both, its shape and size, in the HFL phase. Thus, observation of this feature in dHvA data 
should be good evidence for dominantly local interactions in a system (if $k$-dependence of the self-energy is 
important, the shape of the actual FS will be different from its non- interacting counterpart). In a selective-Mott
metal, on the other hand, it has been proposed that the surface of zeros, defined by $G(k,\omega)=0$~\cite{phillips}, still 
preserves the correct Luttinger count in the strange metal. In our case, we thus expect that the $b$-FS sheet 
(at $\omega=0$ in the orange color plot in Fig.(1)(vi) should still be interpretable as a critical FS without 
the Landau-FL quasiparticles, but that the remnant of $a$-FS sheets must now be understood as loci of 
$G_{aa}(k,\omega=E_{F})=0$ (color plot in Fig.(1)(v)). These two surfaces must together enclose the total 
(original) number of ($a+b$) carriers. In Figs.(1), we show that this is indeed fulfilled in our case: in 
the HFL (upper panel), the $a,b$ FS sheets correspond to poles of $G_{aa,bb}(k,\omega)$ at (a common) $E_{F}$, 
while, in the selective metal [(v)], the $a$-Fermi surface corresponds to the zeros of the selectively Mott localised 
$a$-fermion Green function. The resulting $b$-FS [(vi)] now resembles a critical FS corresponding to loci of branch 
cuts (rather than HFL poles) of $G_{bb}(k,\omega)$ around the nodal $K_{n}=
(\pm\pi/2,\pm\pi/2)$ X-points, describing critical $b$-states, along with patches of zeros of $G_{aa}(k,\omega)$
centered around anti-nodal M points $K_{an}=(\pm\pi,0),(0,\pm\pi)$. Though reminiscent of the pseudogap 
regime of cuprates, these features arise solely from the $d$-wave form factor of the inter-band hybridization in 
our EPAM (there is no dynamical feedback of intersite $d$-wave PG physics into the self-energies in DMFT). 
Importantly, the generalized surface (critical $b$-FS and sheets of zeros of $G_{aa}$ together) lies precisely on
the band Fermi surfaces of the non-interacting problem, and thus encloses the total number of carriers, in 
full accord with the modified Luttinger argument.  As pointed out earlier~\cite{phillips}, the extinction of the Landau quasiparticles and emergence of infra-red singular continuum branch-cut continuua in the OSM phase
reflect emergence of fundamentally new type of collective excitations: these are dubbed ``unparticles'' by Dave {\it et al.}, and described in terms of tomonagons by Anderson~\cite{pwa}.

These results have very interesting implications. If emergence of strange metallicity is linked to an OSMT,
our results offer a way of correlating ARPES and dHvA data as the system is tuned from a HFL into the strange metal
regime: $(i)$ usual LK scaling in dHvA oscillations (with enhanced $m^{*}_{a,b}$) giving a heavily renormalised 
LDA band structure and sharp low-energy peaks (poles) in ARPES lineshapes in the HFL, $(ii)$ anomalous 
($\simeq e^{-T^{1-\alpha}}$) T-dependence of dHvA oscillations, together with anomalously broad infra-red spectral 
functions in ARPES in the strange metal, and $(iii)$ a conventional Luttinger FS in the HFL, partially eaten up by
orbital-selective Mottness in the strange metal. That this is a critical FS is seen from the non-LK scaling 
of dHvA oscillations. In this context, it is important to stress that the $T$-dependent dHvA amplitude ($\chi$) 
can also show a seemingly LK-like form (saturation at low $T$) even in the critical metal~\cite{hartnoll} depending
upon the spectral phase angle in $G_{bb}(k,\omega)$ (see fig.1 in Hartnoll {\it et al.}~\cite{hartnoll}. In fact, even a maximum in $\chi$ maybe seen, providing a 
possible rationale for findings in CeCoIn$_{5}$. Thus, in itself, observation of LK-like scaling of dHvA 
amplitudes alone should not generally be taken as clinching evidence of HFL metallicity. A more comprehensive 
guidance on whether a (Landau)-HFL or strange metallicity obtains in a system requires correlating dHvA and 
ARPES data. Further, anomalous transport (e.g, power-law optical conductivity~\cite{vdM}) also characterises the strange metal,
and this can be used as a further internal self-consistency check for a local critical scenario, since, within DMFT,
the exponents in ARPES lineshapes and optical conductivity should be related to each other (Supplementary 
Information).  Specifically, at least within simplified model hamiltonian contexts, this relation can be used 
to determine the spectral phase angle in $A(k,\omega)$ from the optical phase angle~\cite{pwa}, since the optical 
conductivity is simply the lowest-order bubble diagram involving the full DMFT propagators in the local 
approximation. Having thus extracted the spectral phase angle, the latter can be used in Eq.(1) for the dHvA 
susceptibility to facilitate comparison with experimental data. This procedure should represent an 
internally consistent way to study fermiology in systems where local or quasilocal quantum criticality obtains or 
is expected, e.g, in systems proximate to correlation-driven Mott or partial selective-Mott transitions.

Finally, there is no reason that these features be restricted to the $D=2$ band structure used here. The emergence 
of strange metallicity is tied to a local ``inverse'' lattice OC occuring in the orbital-selective Mott phase, and 
will equally well apply to $D=3$. Further interesting features, related to dynamical effects of non-local 
correlations, could show up in $D=2$: an example is the small (nodal) hole pockets in underdoped cuprates~
\cite{civelli}. Thus, further FS reconstructions occuring as precursors to quasiclassical orders or due 
to strongly fluctuating short-range order may require cluster-DMFT studies, beyond scope of this work.  
However, referring to Fig.(3)b in the local critical metal, we discern specific features pointing toward 
instabilities in the charge and spin sectors, depending on band-filling. Focussing on the $b$-fermion FS 
(orange), we see that enhanced flattening around nodal points, caused by wipe-out of the $k^{2}/2m_{b}^{*}$ 
term in the renormalized dispersion, is propitious for instabilities to AF order if this band is close to being 
half-filled. Furthermore, this feature implies that the leading term in the renormalized ``band'' dispersion will 
now be $\simeq k^{4}$: in real systems with complicated band dispersions with small $\simeq k^{2}$ dispersion
along certain directions in $k$-space, an OSMT like the one we propose may thus wipe-out the $k^{2}$ contribution 
in an anisotropic way.  This maybe relevant to quantum Lifshitz features in CeCu$_{6-x}$Au$_{x}$~\cite{almut} 
or $\beta$-YbAl$_{4}$~\cite{piers}.  We propose that careful dHvA and ARPES studies in these systems may 
offer more comprehensive insight into evolution of fermiology as these systems are driven through their 
anomalous QCPs. In particular, it should shed more light on the extent and influence of quasi-local dynamical 
fluctuations associated with an OSMT on the observed quantum critical features.

Further, the renormalized $a$-band Luttinger surface (or Fermi surface of {\it spinons} if one associates 
fermionic spinons with the fluctuating local moments of the selective-Mott insulating $a$-fermion sector) also 
exhibits pronounced nesting tendency in the nodal directions, allowing for an additional instability to 
local-moment AF order {\it co-existing} with anomalous metallicity. This will further reconstruct the 
Fermi surfaces found here, producing small anti-nodal or nodal pockets.  Which of the two scenarios may 
eventually obtain will depend upon actual system-specific microscopics (e.g, occupation and orbital character of 
$a$ and $b$ bands), and is out of scope of this work. A further possibility is where the leading instability of 
the local strange metal is to (ordered or disordered) valence-bond formation involving the unquenched 
local moemts of the DMFT solution. In this case, the renormalized FS will still be that corresponding to 
the metallic $b$-fermions. However, the fact that the $a$-fermions now form spin-singlet valence bond state(s) 
should drastically reduce the scattering rate (Im$\Sigma_{bb}(\omega)$) and lead to a much more coherent FS. 
This may already have been seen in two-site cluster-DMFT studies for the PAM without $U_{fc}$ and 
a $k$-independent $V_{fc}$~\cite{georges}. DMFT approaches like ours should however suffice, especially in $D=3$, 
in cases where the dHvA FSs in the HFL phase agree (apart from band shifts and mass renormalisations) with LDA 
calculations, since this would imply small ${\bf k}$-dependence of the self-energies. Experimentally, the best 
candidates to test our proposal could be $f$-electron systems like YbRh$_{2}$Si$_{2}$~\cite{steglich}, 
$\beta$-YbAl$_{4}$~\cite{gill-nature}, Ce-based systems~\cite{almut,park} and possibly the very interesting recent finding of anomalous dHvA susceptibility in SmB$_{6}$~\cite{suchitra}, where selective-Mott physics-induced Kondo destruction is 
likely to be implicated in the observed quasi-local quantum critical features. In these systems, a 
clean correlation of dHvA data and anomalous transport remains to be shown. Our proposal can potentially help in this endeavor.

\noindent {\bf Acknowledgements}  MSL wishes to thank the Institute Laue-Langevin, Grenoble for support and Prof. T. Ziman for his kind hospitality whilst this work was begun.  He also thanks Suchitra Sebastian for bringing the fascinating case of SmB$_{6}$ to his notice.

\vspace{1.5cm}

{\bf Supplementary Information}

{\bf Extracting the Spectral Phase Angle}

Here, we show how to extract the spectral phase angle of the one-fermion Green 
function, $G_{bb}(k,\omega)$, and correlate it with the
optical phase angle in the strange metal. We exploit the fact that, in the 
local critical metal, the optical conductivity is just the 
lowest-order bubble diagram composed from the fully dressed one-fermion 
propagators, and vertex corrections rigorously vanish for one-band
systems, and are small enough to be neglected even in the multiband case. 
Referring to earlier DMFT results for our 
extended-PAM, the $b$-fermion propagator in the strange metal has an infra-red 
singular branch-cut structure instead of the heavily renormalized
FL quasiparticle pole structure:

\begin{equation}
Im G_{bb}(i\omega) \simeq \theta(\omega)(i\omega)^{-(1-\alpha)}
\end{equation}

\noindent where $\alpha <1$ results from the ``inverse orthogonality catastrophe'' 
(see main text) arising from strong scattering induced by $U_{fc}$ in the
situation where the one-electron hybridization, $V_{fc}(k)$, is {\it irrelevant} in the renormalization group sense in the underlying impurity
model of DMFT.  Thus, the critical metal has no Landau FL quasiparticles at all.  If one nevertheless chooses to extract a ``quasiparticle'' 
mass and lifetime from such a form by seeking to fit $G_{bb}(k,\omega)$ to a 
hypothetical ``Fermi liquid'' form by identifying an effective
$\omega$-dependent effective mass and quasiparticle scattering rate via 
$m_{qp}(\omega)/m_{b}=z^{-1}(\omega)$ and $\tau_{qp}^{-1}=$Im$\Sigma_{b}(\omega)$ (where $\Sigma_{b}(\omega)$ is the one-particle irreducible self energy).
 In the local critical metal, the spectral phase angle is thus $\phi=(1-\alpha)\pi/2$.
  Within DMFT, the optical conductivity is directly computable as a ``simple'' bubble diagram composed of the
fully renormalized DMFT propagators. Using the form of $G_{bb}(\omega)$ above, 
one infers from scaling arguments that $\sigma(\omega) \simeq (i\omega)^{-(1-\eta)}$: 
explicit DMFT results for the EPAM indeed bear 
this out in the selective-Mott phase. The {\it optical} phase angle, first 
measured by van der Marel {\it et al.}~\cite{vdM}, is now $(1-\eta)\pi/2$.
 Since the exponents in $G_{bb}(\omega)$ and $\sigma(\omega)$ are 
obviously closely related within DMFT, extracting the optical phase angle 
allows an estimate of the spectral phase angle $\phi$. It is this spectral 
phase angle which will determine the actual $T$-dependence of the 
dHvA susceptibility, as done by Hartnoll {\it et al}, as alluded to in the main text.
In particular, it can show either a monotonically increasing form, a LK-like $T$-independent form, or
a finite-$T$ maximum, dependent upon the $\phi$ from the above procedure.  This procedure can thus achieve internal self-consistency between dHvA, ARPES and anomalous transport in strange metals. 

\end{document}